\documentclass[aps,prb,preprint,floatfix]{revtex4-1}
\usepackage{amsmath,amssymb,epsfig}

\begin{document}
  

\title{Exact ground state for the four-electron problem in a 2D finite 
honeycomb lattice}

\author{R\'eka~Trencs\'enyi$^{a}$, Konstantin~Glukhov$^{b}$, 
and Zsolt~Gul\'acsi$^{c}$}
\address{
$^{(a)}$ Institute for Nuclear Research, Hungarian Academy of Sciences, 
H-4026 Debrecen, Bem ter 18/c \\
$^{(b)}$ Institute for Solid State Physics and Chemistry, Uzhgorod National
University, Voloshyn Street 54, Uzhgorod 88000, Ukraine\\
$^{(c)}$ Department of Theoretical Physics, University of Debrecen, 
H-4010 Debrecen, Hungary}

\date{March 5, 2014}

\begin{abstract}
Working in a subspace with dimensionality much smaller than the dimension 
of the full Hilbert space, we deduce
exact 4-particle ground states in 2D samples containing hexagonal 
repeat units and described by Hubbard type of models. 
The procedure identifies first
a small subspace ${\cal{S}}$ in which the ground state $|\Psi_g\rangle$ 
is placed, than deduces $|\Psi_g\rangle$ by exact diagonalization in
${\cal{S}}$. The small subspace is obtained by the repeated application of the
Hamiltonian $\hat H$ on a carefully chosen starting wave vector describing
the most interacting particle configuration, and the wave
vectors resulting from the application of $\hat H$, till the obtained system
of equations closes in itself. The procedure which can be applied in principle 
at fixed but arbitrary system size and number of particles,
is interesting by its own since
provides exact information for the numerical approximation techniques which 
use a similar strategy, but apply non-complete basis for ${\cal{S}}$.
The diagonalization inside ${\cal{S}}$ provides an incomplete image about the
low lying part of the excitation spectrum, but provides the exact  
$|\Psi_g\rangle$. 
Once the exact ground state is obtained, its properties can be easily analyzed.
The $|\Psi_g\rangle$  is found always as a singlet state whose energy, 
interestingly, saturates in the $U \to \infty$ limit. The unapproximated 
results show that 
the emergence probabilities of different particle configurations in the 
ground state present ``Zittern'' (trembling) characteristics which are absent 
in 2D square Hubbard systems. Consequently, the manifestation of the local 
Coulomb repulsion in 2D square and honeycomb types of systems presents 
differences, which can be a real source in the differences in 
the many-body behavior.  
\end{abstract}
\maketitle


\section{Introduction}

Systems containing few fermions are interesting by their own. From one side,
they are analyzed because of their {\it in principle} importance, as for
example providing lower bounds to the ground state energy of more complicated
systems containing identical, but arbitrary high number of particles $N$
\cite{Intr1}, lead to potentially valuable and non-perturbative
information regarding the 
many-body behavior \cite{Intr2,Intr3} as demonstrated by \cite{Intr4,Intr5},
or directly relate to basic principles of quantum theory, as for example
non-locality derived from entanglement in the four-particle case \cite{Intr5a}.
From the other hand, experimental developments of the last years allow to
confine small number of atoms in a trap and address directly their quantum
state \cite{Intr6,Intr7,Intr8}. On this background, the few-fermion states
have been intensively studied with focus on different aspects, as for example
emergence possibilities of inhomogeneous condensate \cite{Intr9},
effect of the Coulomb interaction \cite{Intr10}, or study of bound states
\cite{Intr11}. The investigations start in fact from the two-particle level 
\cite{Intr2,Intr3,Intr12}, the three-particle cases abound for example
in the study of the behavior in harmonic trap \cite{Intr13}, development of 
effective theories \cite{Intr14}, study of the Efimov effect \cite{Intr15},
characterization of contact parameters in 2D \cite{Intr15x}, behavior in 1D
trap \cite{Intr15y},
or in describing quantum dot systems \cite{Intr16}. Besides experimental
observations \cite{Intr5a,Intr11}, theoretical 
investigations for the four-particle cases are also present, mostly by 
numerical descriptions using exact diagonalization \cite{Intr9}, or effective 
theories \cite{Intr10}.
However, connected to, and originating from the search for 
techniques leading to 
non-approximated results for non-integrable systems in 
one \cite{Intr4,Intr5,I10b,I10c}, two \cite{II10a}, and
three \cite{III10a} dimensions, also exact results are present for the
four particle problem in the 2D square
Hubbard case \cite{I1}, or Hubbard ladders \cite{I1a}.

In this paper we concentrate on 2D systems built up from periodic
hexagonal repeat units,
as encountered in honeycomb or graphene type of lattices, being interested to
deduce valuable good quality information relating the effects of the 
interaction on the many-body behavior. One knows that in such systems, because
of the coupling constant value, nor perturbative expansions, nor strong
coupling theories are properly justified \cite{Intr017}, but in the same
time, in the study of graphene type of materials, a strong need of 
non-perturbative input is present \cite{Intr17}. Furthermore, controversies
relating the differences in the caused effects of the interaction in 2D
systems with square and hexagonal repeat units \cite{Intr18,Intr19,Intr20}
also demand good quality input relating interaction driven many-body effects
in these systems. 

Starting from these requirements, in the present paper we present exact 
four-particle ground states for 2D honeycomb  samples with periodic boundary
conditions taken in both directions. The method \cite{I1} is based on deducing 
a small subspace ${\cal{S}}$ containing the $|\Psi_g\rangle$ ground state wave 
function in exact terms, followed by the non-approximated calculation of 
different ground state characteristics. The technique itself starts from a
wave vector $|1\rangle$ containing the most interacting particle configuration  
translated to each site of the lattice and added. The Hamiltonian $\hat H$
acting on the vector $|1\rangle$ generates vectors $|i\rangle$ with similar
properties (i.e. a local particle configuration taken at each site and added),
and the linear system of equations closes in itself
\begin{eqnarray}
\hat H |i\rangle=\sum_j \alpha_{j,i} |j\rangle,
\label{Eq0}
\end{eqnarray}
after a number of steps much
less than the dimensionality of the full Hilbert space. This generates the
subspace ${\cal{S}}$ containing the exact ground state. The method in 
principle can be applied independent on the system size and independent on the
fixed number of identical particles inside the system. The results are 
interesting not only because provide in 2D honeycomb systems 
an exact four particle ground state which
has its fingerprint in more complicated ground states holding an arbitrary 
high number of particles N \cite{Intr1}. The results are also important
because in the last years,  procedures generating limited functional
spaces based on the system in (\ref{Eq0}) cut after a given number of 
steps (i.e. using an incomplete ${\cal{S}}$ basis), started to be used in 
different approximations and numerical approaches
\cite{Intr20a,Intr20b}, for which, exact results could provide an important
insight.

Turning back to the deduced exact four particle ground state,
the results show that the ground state of the system
is a total spin singlet $S=0$ state which has a ground state energy $E_g$ 
that interestingly saturates at a finite value, for increasing interaction
strength, in the $U \to \infty$ limit.  Furthermore, $|\Psi_g\rangle$ has a
special property not present in the 2D square Hubbard system, namely that the
emergence probability of different particle configurations in the ground state
wave function present trembling (``Zittern'') in function of $U$. Because of 
this, small modifications in the interaction strength in 2D honeycomb systems 
could cause main changes in the many-body behavior, which 
underlines differences in the many-body effects of the interaction 
in 2D lattices with hexagonal and square repeat units, 

The remaining part of the paper is organized as follows: Section II. presents
the studied system, Section III. describes the method and the deduced 
four-particle ground state, Section IV. describes the observed properties of 
the ground state, Section V. contains the discussions and 
summary, while finally, two appendices A, and B containing mathematical
details close the presentation.

\section{The studied system} %

In order to analyze the four-electron problem in a graphene type of system,
one takes a two dimensional array of periodically displaced hexagons with 
equivalent sites, cutting from this $m_h=n_{h,h} \times n_{h,v}$ neighboring 
hexagons
and treating them with a Hubbard type of model and periodic boundary conditions.
A such kind of system becomes in fact a torus with $n_{h,h}$ hexagons displaced
along the ring of the torus (i.e. along the toroidal direction) and $n_{h,v}$ hexagons 
along the poloidal direction, providing the thickness (i.e. the ring
circumference) of the sectioned torus ring. 
\begin{figure} [h]                                                         
\centerline{\includegraphics[width=10cm,height=6.5cm]{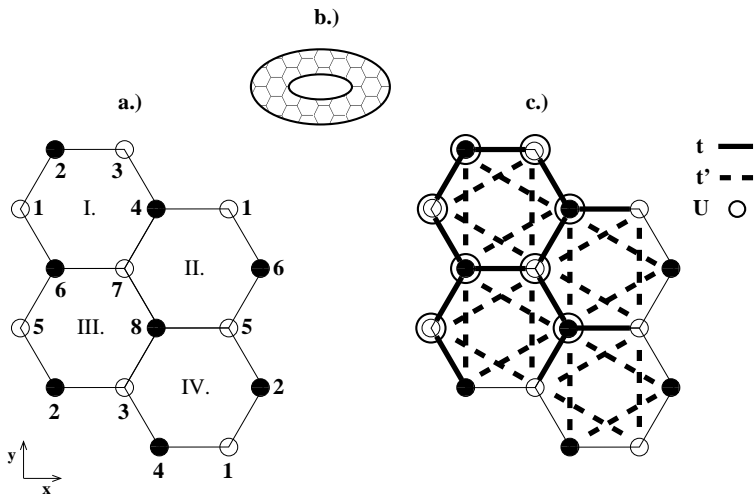}} 
\caption{a) The studied system. The hexagons, (sites) are labeled by the 
index $J=I,II,III,IV$, ($n=1,2,...,8$), and sublattices are indicated by 
black and white dots. b) The torus-like shape of the system taken with
periodic boundary conditions. c) The used Hamiltonian parameters.}      
\end{figure}                                                               
The smallest nontrivial system 
of this type, which retains main properties of the interacting four-electron 
problem, in the studied situation has $n_{h,h}=n_{h,v}=2$, hence four constituent 
hexagons with $N_{\Lambda}=8$ different sites, and the corresponding four-electron 
problem in the singlet case has a 784 dimensional Hilbert space. Being the 
easiest to treat, we analyze below this case, but the procedure we apply is 
the same for arbitrary $m_h$. The system is presented in Fig.1, while the 
Hamiltonian $\hat H=\hat H_{kin}+\hat H_U$, $\hat H_{kin}=\hat T_1 + \hat T_2$ is
given by  
\begin{eqnarray}
\hat T_1 &=& \sum_{\sigma} [
t \: \hat c^{\dagger}_{1,\sigma} \hat c_{2,\sigma} +
t \: \hat c^{\dagger}_{1,\sigma} \hat c_{4,\sigma} +
t \: \hat c^{\dagger}_{1,\sigma} \hat c_{6,\sigma} +
t \: \hat c^{\dagger}_{2,\sigma} \hat c_{3,\sigma} +
t \: \hat c^{\dagger}_{2,\sigma} \hat c_{5,\sigma} +
t \: \hat c^{\dagger}_{3,\sigma} \hat c_{4,\sigma}
\nonumber\\
&+&
t \: \hat c^{\dagger}_{3,\sigma} \hat c_{8,\sigma} +
t \: \hat c^{\dagger}_{4,\sigma} \hat c_{7,\sigma} +
t \: \hat c^{\dagger}_{5,\sigma} \hat c_{6,\sigma} +
t \: \hat c^{\dagger}_{5,\sigma} \hat c_{8,\sigma} +
t \: \hat c^{\dagger}_{6,\sigma} \hat c_{7,\sigma} +
t \: \hat c^{\dagger}_{7,\sigma} \hat c_{8,\sigma} + H.c.],
\nonumber\\
\hat T_2 &=& \sum_{\sigma} [
t'_{I} \: \hat c^{\dagger}_{1,\sigma} \hat c_{3,\sigma} +
t'_{IV} \: \hat c^{\dagger}_{1,\sigma} \hat c_{3,\sigma} +
t'_{II} \: \hat c^{\dagger}_{1,\sigma} \hat c_{5,\sigma} +
t'_{IV} \: \hat c^{\dagger}_{1,\sigma} \hat c_{5,\sigma} +
t'_{I} \: \hat c^{\dagger}_{1,\sigma} \hat c_{7,\sigma} +
t'_{II} \: \hat c^{\dagger}_{1,\sigma} \hat c_{7,\sigma}
\nonumber\\
&+&
t'_{I} \: \hat c^{\dagger}_{2,\sigma} \hat c_{4,\sigma} +
t'_{IV} \: \hat c^{\dagger}_{2,\sigma} \hat c_{4,\sigma} +
t'_{I} \: \hat c^{\dagger}_{2,\sigma} \hat c_{6,\sigma} +
t'_{III} \: \hat c^{\dagger}_{2,\sigma} \hat c_{6,\sigma} +
t'_{III} \: \hat c^{\dagger}_{2,\sigma} \hat c_{8,\sigma} +
t'_{IV} \: \hat c^{\dagger}_{2,\sigma} \hat c_{8,\sigma}
\nonumber\\
&+&
t'_{III} \: \hat c^{\dagger}_{3,\sigma} \hat c_{5,\sigma} +
t'_{IV} \: \hat c^{\dagger}_{3,\sigma} \hat c_{5,\sigma} +
t'_{I} \: \hat c^{\dagger}_{3,\sigma} \hat c_{7,\sigma} +
t'_{III} \: \hat c^{\dagger}_{3,\sigma} \hat c_{7,\sigma} +
t'_{I} \: \hat c^{\dagger}_{4,\sigma} \hat c_{6,\sigma} +
t'_{II} \: \hat c^{\dagger}_{4,\sigma} \hat c_{6,\sigma}
\nonumber\\
&+&
t'_{II} \: \hat c^{\dagger}_{4,\sigma} \hat c_{8,\sigma} +
t'_{IV} \: \hat c^{\dagger}_{4,\sigma} \hat c_{8,\sigma} +
t'_{II} \: \hat c^{\dagger}_{5,\sigma} \hat c_{7,\sigma} +
t'_{III} \: \hat c^{\dagger}_{5,\sigma} \hat c_{7,\sigma} +
t'_{II} \: \hat c^{\dagger}_{6,\sigma} \hat c_{8,\sigma} +
t'_{III} \: \hat c^{\dagger}_{6,\sigma} \hat c_{8,\sigma} 
\nonumber\\
&+& H.c.],
\nonumber\\
\hat H_U &=&
U \: \hat n_{1,\uparrow} \hat n_{1,\downarrow} +
U \: \hat n_{2,\uparrow} \hat n_{2,\downarrow} +
U \: \hat n_{3,\uparrow} \hat n_{3,\downarrow} +
U \: \hat n_{4,\uparrow} \hat n_{4,\downarrow} +
U \: \hat n_{5,\uparrow} \hat n_{5,\downarrow} +
U \: \hat n_{6,\uparrow} \hat n_{6,\downarrow} 
\nonumber\\
&+&
U \: \hat n_{7,\uparrow} \hat n_{7,\downarrow} +
U \: \hat n_{8,\uparrow} \hat n_{8,\downarrow},
\label{Eq1}
\end{eqnarray}
where $\hat c^{\dagger}_{i,\sigma}$ creates an electron with spin projection 
$\sigma$ on site $i$,  $U \geq 0$ characterizes the local Coulomb repulsion,
t represents the nearest neighbor hopping matrix element, and $t'_J=t'$ is the 
next nearest neighbor hopping matrix element inside the hexagon $J=I,II,III,IV$.

\section{The applied procedure} %

\subsection{The basic strategy of the method}

The technique we apply, which has never been used in the study of 2D materials 
with hexagonal repeat units, has been described in details in Ref.\cite{I1},
where it has been successfully utilized in deriving the four-electron ground 
state for finite 2D Hubbard model on a square lattice. The method works for 
singlet ground states $|\Psi_g\rangle$ provided by an arbitrary even number of
electrons, N, whose Hilbert space is ${\cal{H}}$. The procedure itself is 
based on the identification of a small subspace ${\cal{S}}$ in which the 
ground state is placed giving rise to the exact, explicit and handable   
expression of the multielectronic ground state wave function. 
For example, in case of the 2D square Hubbard system analyzed in 
Ref. \cite{I1}, it was shown that for $N=4$ electrons and 
$N_{\Lambda}=4\times 4= 16$ sites, for 
which the Hilbert space dimensionality is $Dim({\cal{H}})=14400$,
the subspace ${\cal{S}}$ containing $|\Psi_g\rangle$ has only the dimension
$Dim({\cal{S}})=85$. Hence, in the process of deducing $|\Psi_g\rangle$ for the
square system in Ref.\cite{I1}, working in ${\cal{S}}$ instead of ${\cal{H}}$, 
one has a 170 times of dimensionality reduction (i.e. two orders of magnitude).

The method constructs the basis vectors of ${\cal{S}}$ based on the following 
strategy: 
i) the most interacting particle configuration is part of $|\Psi_g\rangle$, 
and ii) being a translational invariant system, the most interacting particle
configuration is equally present with the same weight around all lattice 
sites. Starting from i),ii), the first base vector $|1\rangle$  of  
${\cal{S}}$ is constructed by taking the most interacting particle 
configuration, translating it to all lattice sites, and adding together all 
contributions. Once $|1\rangle$ exists, the other base vectors are obtained by
the action of the Hamiltonian. This is based on the fact that iii) if a wave
vector $|j\rangle$ was such constructed that a particle configuration was
translated to all lattice sites and all such obtained contributions were added,
than by the action of the Hamiltonian on $|j\rangle$, the obtained new wave
vectors $|j'\rangle$ have similar properties, but related to different particle 
configurations. Consequently, $\hat H |1\rangle$ produces new $|j'\rangle$
vectors, which, if linearly independent, will be considered new base vectors of
${\cal{S}}$, i.e. $|2\rangle$, $|3\rangle$, etc. Similarly, $\hat H |2\rangle$,
$\hat H |3\rangle$, etc. give rise to new base vectors. The procedure is 
applied till the set of base vectors $\{|1\rangle, |2\rangle,...
|N_{\cal{S}}\rangle$ closes in itself. 

In order to clarify the used strategy, let us enumerate 
first the possible particle configurations which can appear 
in the present case for 4 electrons in a singlet state. One has
three possibilities, namely a) two double occupied sites, b) one
double occupied site and two electrons with opposite spin on two other
different sites, and c) two electrons with spin up and two electrons with spin 
down, all on different sites. These three possibilities are graphically 
presented in Fig.2, where a black dot on site ${\bf i}$ represents a double 
occupancy at the site ${\bf i}$ (see Fig.2.a), a dashed line connecting two
different sites ${\bf j} \ne {\bf k}$ represents two electrons, one with spin 
$\sigma$ at the site ${\bf j}$ and one with spin $-\sigma$ at the site 
${\bf k}$, where $\sigma$ is arbitrary (see Fig.2.b), and finally, a full line
connecting two different sites ${\bf j} \ne {\bf i}$ represents two electrons
placed with the same spin $\sigma$ on the sites ${\bf j}$ and  ${\bf i}$,
$\sigma$ is arbitrary (see Fig.2.c).

\begin{figure} [h]                                                  
\centerline{\includegraphics[width=9cm,height=3.3cm]{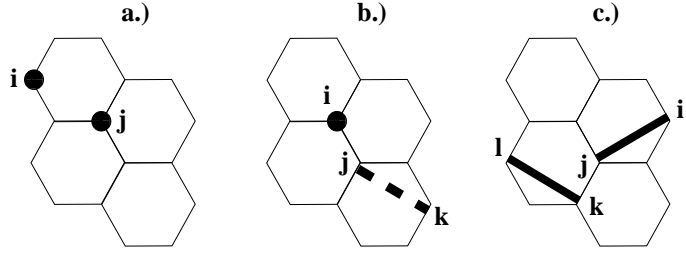}} 
\caption{The three possible types of electron       
states in the system. Black dot at the site ${\bf i}$ means a double 
occupancy at ${\bf i}$, a dashed line connecting two
different sites ${\bf j} \ne {\bf k}$ represents two electrons, one with spin 
$\sigma$ at the site ${\bf j}$ and one with spin $-\sigma$ at the site 
${\bf k}$, where $\sigma$ is arbitrary, and finally, a full line
connecting two different sites ${\bf j} \ne {\bf i}$ represents two electrons
placed with the same spin $\sigma$ on the sites ${\bf j}$ and  ${\bf i}$, 
$\sigma$ is arbitrary. The plots a), b), and c) represent the three different 
possibilities that can appear for four electrons in a singlet state.}                           
\end{figure}                                                        

The mathematical expressions of the normalized wave vectors connected to the 
graphical presentations in Fig.2 are as follows: Fig.2.a means
\begin{eqnarray}
|\psi_a(i,j) \rangle = 
\big(\hat c_{i,\uparrow}^{\dagger} \hat c_{i,\downarrow}^{\dagger}\big)
\big(\hat c_{j,\uparrow}^{\dagger} \hat c_{j,\downarrow}^{\dagger}\big) |0 \rangle,
\label{Eq2}
\end{eqnarray}
where $|0 \rangle$ represents the bare vacuum, and one has 
$i \neq j$ and $i<j$.

The mathematical expression connected to Fig.2.b is
\begin{eqnarray}
|\psi_b(i;j,k) \rangle = \frac{1}{\sqrt{2}}
\big(\hat c_{i,\uparrow}^{\dagger} \hat c_{i,\downarrow}^{\dagger}\big)
\big[\big(\hat c_{j,\uparrow}^{\dagger} \hat c_{k,\downarrow}^{\dagger}\big) +
\big(\hat c_{k,\uparrow}^{\dagger} \hat c_{j,\downarrow}^{\dagger}\big)\big] |0 \rangle,
\label{Eq3}
\end{eqnarray}
where $i \neq j$, $i \neq k$ and $j \neq k$ is required.

Finally, the mathematical meaning of Fig.2.c is given by
\begin{eqnarray}
|\psi_c(j,i;l,k) \rangle = \frac{1}{\sqrt{2}}
\big[\big(\hat c_{j,\uparrow}^{\dagger} \hat c_{i,\uparrow}^{\dagger}\big)
\big(\hat c_{l,\downarrow}^{\dagger} \hat c_{k,\downarrow}^{\dagger}\big) +
\big(\hat c_{j,\downarrow}^{\dagger} \hat c_{i,\downarrow}^{\dagger}\big)
\big(\hat c_{l,\uparrow}^{\dagger} \hat c_{k,\uparrow}^{\dagger}\big)\big] |0 \rangle,
\label{Eq4}
\end{eqnarray}
where $i \neq j \neq k \neq l$, together with $j>i$ and $l>k$ must be satisfied.

A given particle configuration is an arbitrary four-electron state contained in
one of the vectors presented in Eqs.(\ref{Eq2},\ref{Eq3},\ref{Eq4}). The most
interacting particle configuration includes two double occupancies placed at
nearest neighbor sites.

\subsection{The application of the method}

The application of the method consists basically of three steps, namely:
a) the construction of a starting base vector, b) the application of the 
Hamiltonian on the starting wave vector and collecting the resultant base 
vectors
describing also resultant configurations placed at different sites and added
together, c) further application of the Hamiltonian on the resultant base 
vectors till the system closes (i.e. new resultant linearly independent 
vectors no more appear).
This happens after a number of steps $N_{\cal{s}}$ (i.e. a number of equations 
$N_{\cal{s}}$), which is usually orders of magnitude smaller than 
$Dim({\cal{H}})$. In the present case $N_{\cal{s}}=Dim({\cal{S}})=70$. Below we 
describe the steps a),b),c) in details.

For the first step, a) we take into consideration the basic starting points of 
the method. Consequently, one starts with a most 
interacting configuration ($|\psi_a(2,3) \rangle$) and writes it on all 
(sublattice) sites
($|\psi_a(6,7) \rangle, |\psi_a(5,8) \rangle, |\psi_a(1,4) \rangle$). Since all
these configurations must appear with the same weight, we add all these 
contributions, normalize the sum and obtain the starting base vector of
${\cal{S}}$ as
\begin{eqnarray}
|1\rangle = \frac{1}{2}(|\psi_a(2,3) \rangle + |\psi_a(6,7) \rangle +
|\psi_a(5,8) \rangle + |\psi_a(1,4) \rangle),
\label{Eq5}
\end{eqnarray}
which is represented in graphical form in the first position of Fig.3. Note
that one has in the studied sample four sublattice sites, so $|1\rangle$ must 
have four components.

For the step b) we simply apply the Hamiltonian on $|1\rangle$, obtaining
\begin{eqnarray}
\hat H |1\rangle &=& 2U |1\rangle + 2t |8\rangle + 2t |10\rangle + 4t' 
|17\rangle + 4t' |18\rangle + 4t' |22\rangle,
\label{Eq6}
\end{eqnarray}
where the new resultant linearly independent base vectors denoted by 
$|8 \rangle, |10 \rangle, 
|17 \rangle, |18 \rangle, |22 \rangle$ can be seen in Figs.3-4. We note that 
because of the clarity of the presentation, the numbering of the base vectors 
not follows the order of appearance, but the 
constituent type. The mathematical expressions corresponding to the graphical
representations in Figs.3-8 are simple: for a given vector take every plotted 
contribution, write them in mathematical form according to the rules described 
in Eqs.(\ref{Eq2},\ref{Eq3},\ref{Eq4}), add all contributions together and 
finally, normalize the sum.

Now the step c) follows: one applies the Hamiltonian on all new resultant base
vectors, obtaining
\begin{eqnarray}
\hat H |8\rangle &=& 2t |1\rangle + 2t |4\rangle + U |8\rangle + 2t' 
|10\rangle + 2t' |14\rangle + t |22\rangle + t |25\rangle + 
2t' |27\rangle + 2t' |28\rangle 
\nonumber\\
&-& 2t |37\rangle - 2t' |42\rangle - 2t' |43\rangle - 2t' |47\rangle + 
t |52\rangle - 2t |53\rangle + t |57\rangle - 2t' |65\rangle,
\nonumber\\
\hat H |10\rangle &=& 2t |1\rangle + 2t |5\rangle + 2t' |8\rangle + 
U |10\rangle + 2t' |15\rangle + t |22\rangle + t |25\rangle + 
2t' |26\rangle + 2t' |28\rangle 
\nonumber\\
&-& 2t |38\rangle + 2t' |41\rangle - 2t' |44\rangle + 2t' |47\rangle + 
t |52\rangle + 2t |54\rangle + t |57\rangle + 2t' |66\rangle,
\nonumber\\
\hat H |17\rangle &=& 4t' |1\rangle + 4t' |3\rangle + U |17\rangle + 2t' 
|18\rangle + 2t' |19\rangle + 2t' |20\rangle + 
2t' |22\rangle + t |26\rangle + t |28\rangle 
\nonumber\\
&-& 4t' |29\rangle + 2t' |32\rangle + 2t' |34\rangle + 4t' |37\rangle + 
t |41\rangle + t |47\rangle - 2t' |50\rangle - 2t' |52\rangle,
\nonumber\\
\hat H |18\rangle &=& 4t' |1\rangle + 4t' |2\rangle + 2t' |17\rangle + U 
|18\rangle + 2t' |19\rangle + 2t' |21\rangle + 
2t' |22\rangle + t |27\rangle + t |28\rangle 
\nonumber\\
&+& 4t' |30\rangle + 2t' |31\rangle + 2t' |34\rangle + 4t' |38\rangle - 
t |42\rangle - t |47\rangle + 2t' |51\rangle - 2t' |52\rangle,
\nonumber\\
\hat H |22\rangle &=& 4t' |1\rangle + 4t' |7\rangle + t |8\rangle + t 
|10\rangle + t |14\rangle + t |15\rangle + 2t' |17\rangle + 
2t' |18\rangle + 2t' |20\rangle 
\nonumber\\
&+& 2t' |21\rangle + U |22\rangle + 2t' |31\rangle + 2t' |32\rangle - 
4t' |33\rangle - t |43\rangle - t |44\rangle - 2t' |50\rangle 
\nonumber\\
&+& 2t' |51\rangle - t |65\rangle + t |66\rangle + 4t' |68\rangle,
\label{Eq7}
\end{eqnarray}
where the new resulting base vectors can be seen in Figs.(3-8). 
Repeating the Hamiltonian action on the newly resulting vectors, the system 
closes after 70 steps (i.e. after 70 equations). The whole system of equations 
is presented in Appendix A, and all emerging contributions are depicted in 
Figs.(3-8).
We note that the normalized and orthogonal vectors $|n\rangle$, $n=1,2,...,70$, 
represent the base vectors of the subspace ${\cal{S}}$.

\newpage

\begin{figure} [h]                                                      %
\centerline{\includegraphics[width=12cm,height=18cm]{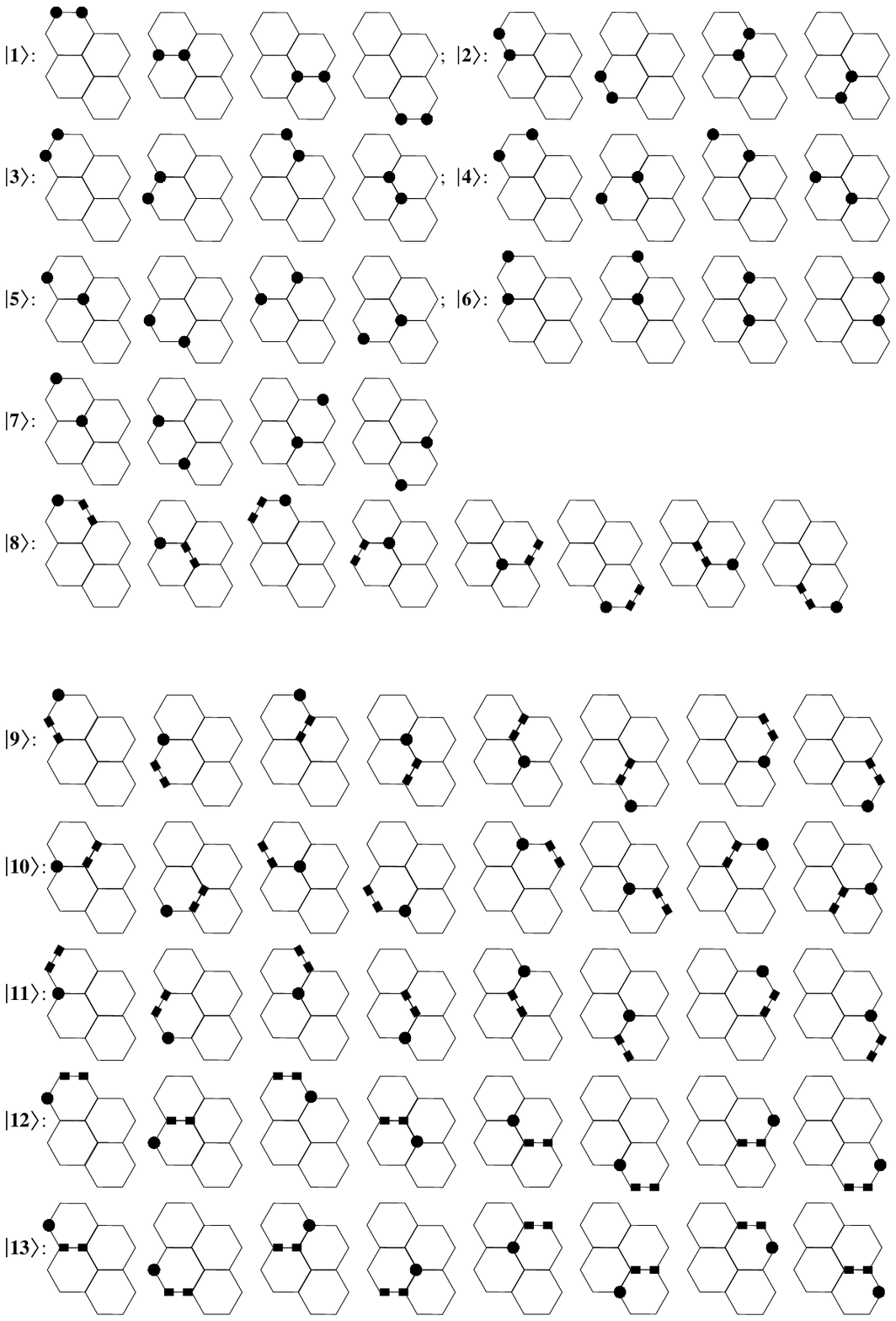}} %
\caption{The structure of the base vectors $|1\rangle-|13\rangle$ of    %
${\cal{S}}$. The black dot at a given site means a double occupancy at  %
that site, while a dashed line connecting two different sites           %
${\bf i} \ne {\bf j}$ represents two electrons, one with spin $\sigma$  %
at the site ${\bf i}$ and one with spin $-\sigma$ at the site           %
${\bf j}$, where $\sigma$ is arbitrary.}                                %
\end{figure}                                                            %

\newpage

\begin{figure} [h]                                                      %
\centerline{\includegraphics[width=12cm,height=18cm]{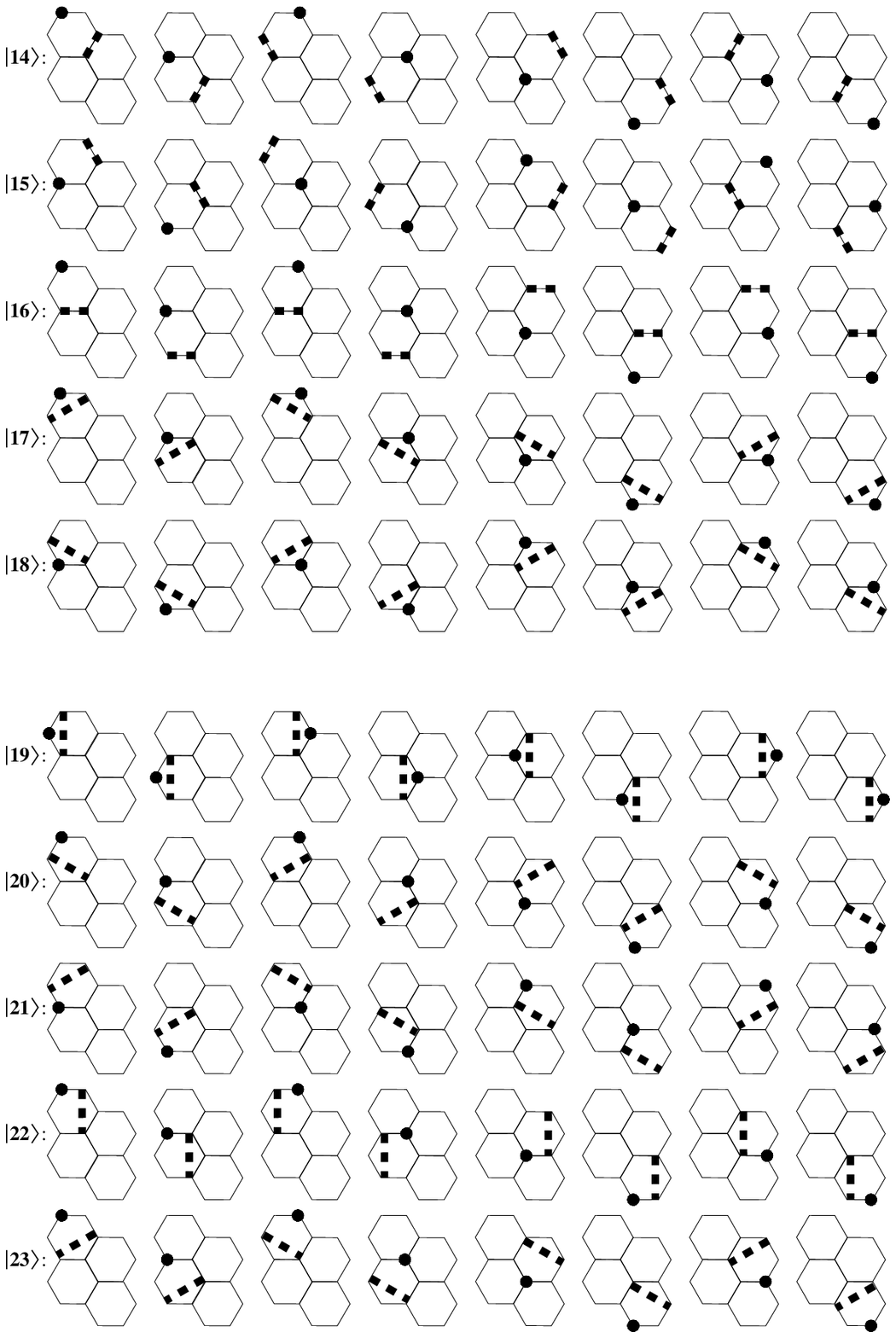}} %
\caption{The structure of the base vectors $|14\rangle-|23\rangle$ of   %
${\cal{S}}$. The meaning of the notations is as in Fig.3.}              %
\end{figure}                                                            %

\newpage

\begin{figure} [h]                                                      %
\centerline{\includegraphics[width=12cm,height=18cm]{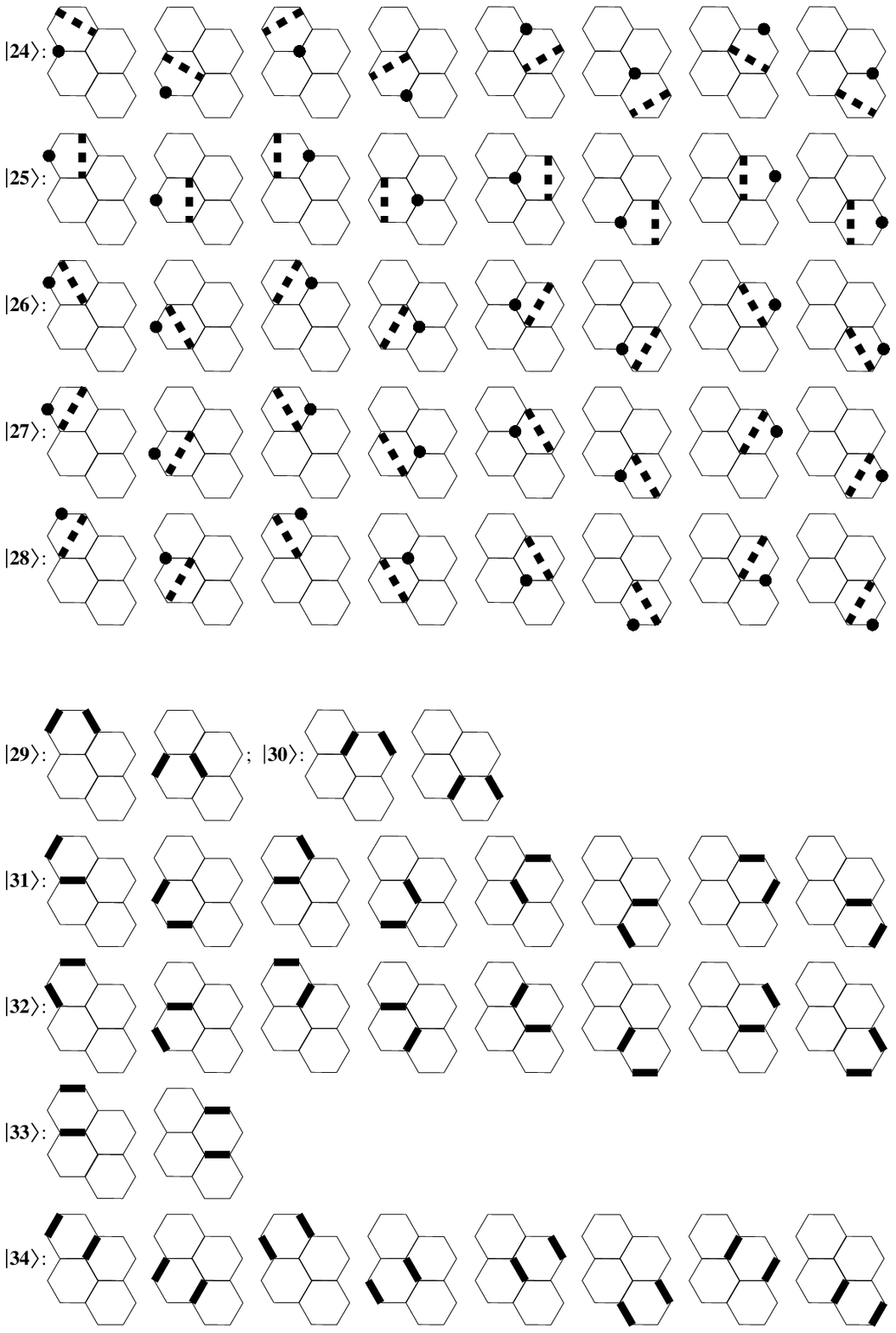}} %
\caption{The structure of the base vectors $|24\rangle-|34\rangle$ of   %
${\cal{S}}$. Up to the vector $|28\rangle$ the meaning of the notations %
is as in Fig.3. Starting from the vector $|29\rangle$, the full line    %
connecting two different sites ${\bf i} \ne {\bf j}$ represents two     %
electrons placed with the same spin $\sigma$ on the sites ${\bf i}$ and %
${\bf j}$, $\sigma$ being arbitrary.}                                   %
\end{figure}                                                            %

\newpage

\begin{figure} [h]                                                      %
\centerline{\includegraphics[width=12cm,height=18cm]{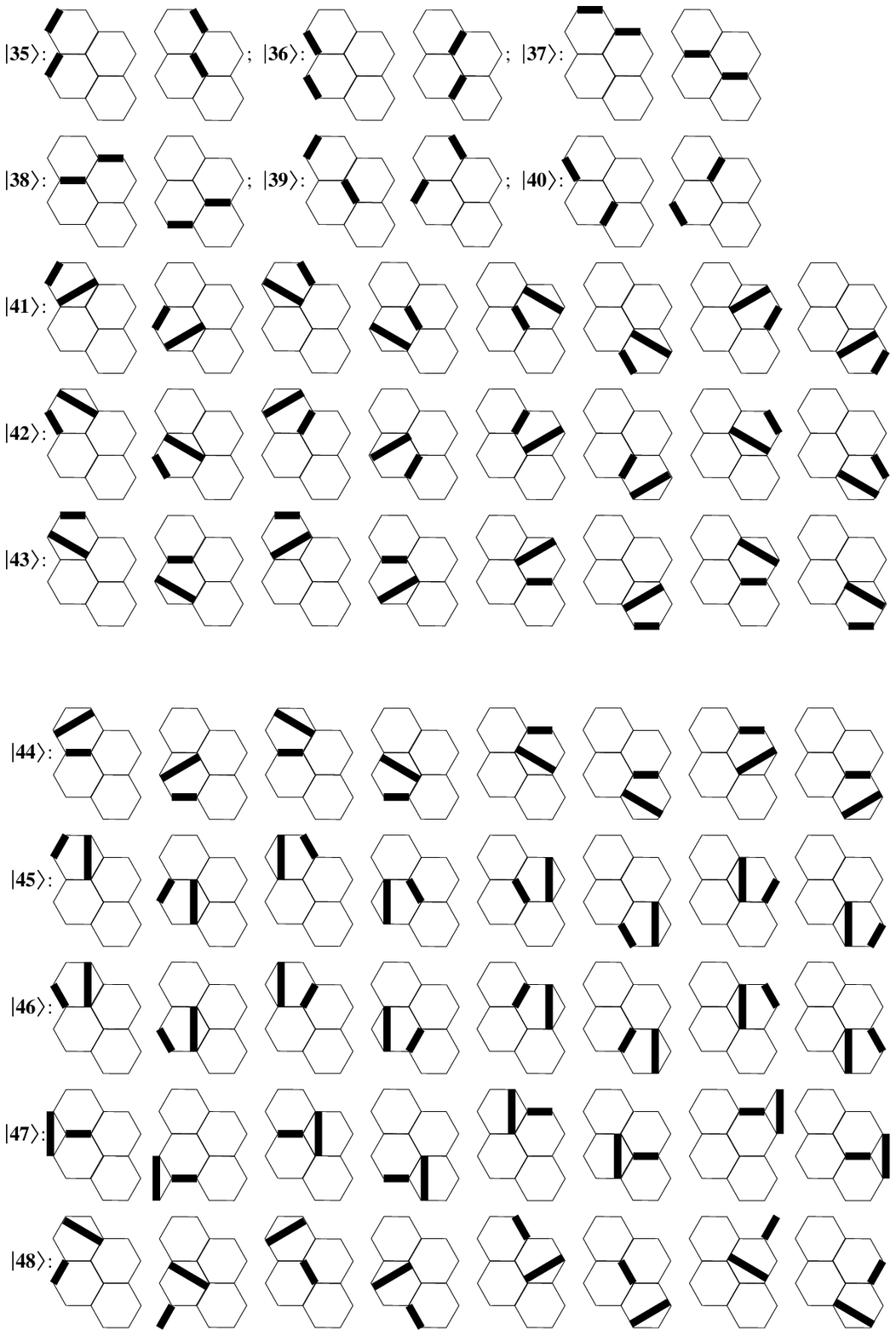}} %
\caption{The structure of the base vectors $|35\rangle-|48\rangle$ of   %
${\cal{S}}$. The meaning of the notations is as in Figs.(3-5).}         %
\end{figure}                                                            %

\newpage

\begin{figure} [h]                                                        %
\centerline{\includegraphics[width=12cm,height=18.5cm]{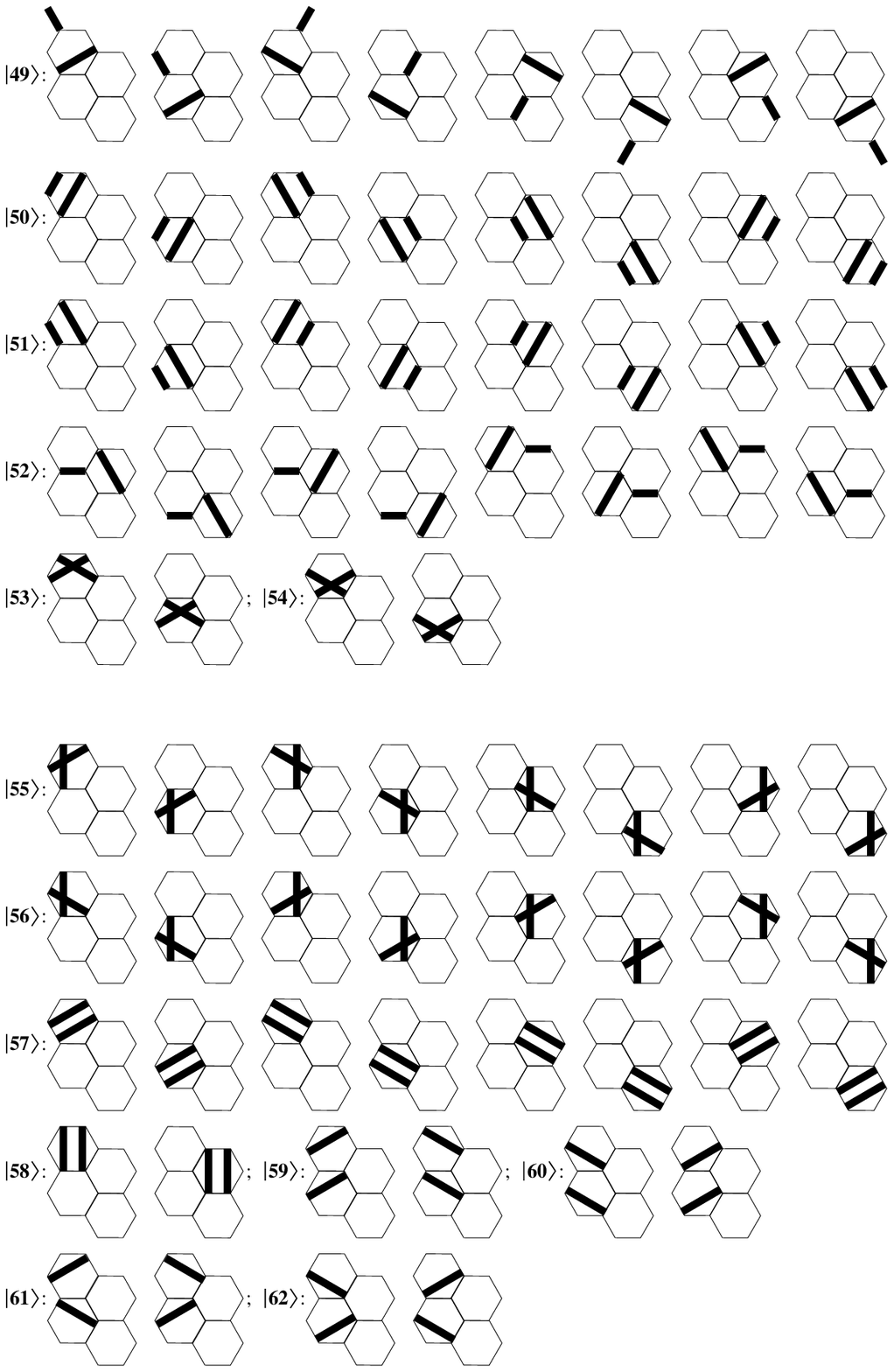}} %
\caption{The structure of the base vectors $|49\rangle-|62\rangle$ of     %
${\cal{S}}$. The meaning of the notations is as in Figs.(3-5).}           %
\end{figure}                                                              %

\newpage

\begin{figure} [h]                                                       %
\centerline{\includegraphics[width=12cm,height=8.5cm]{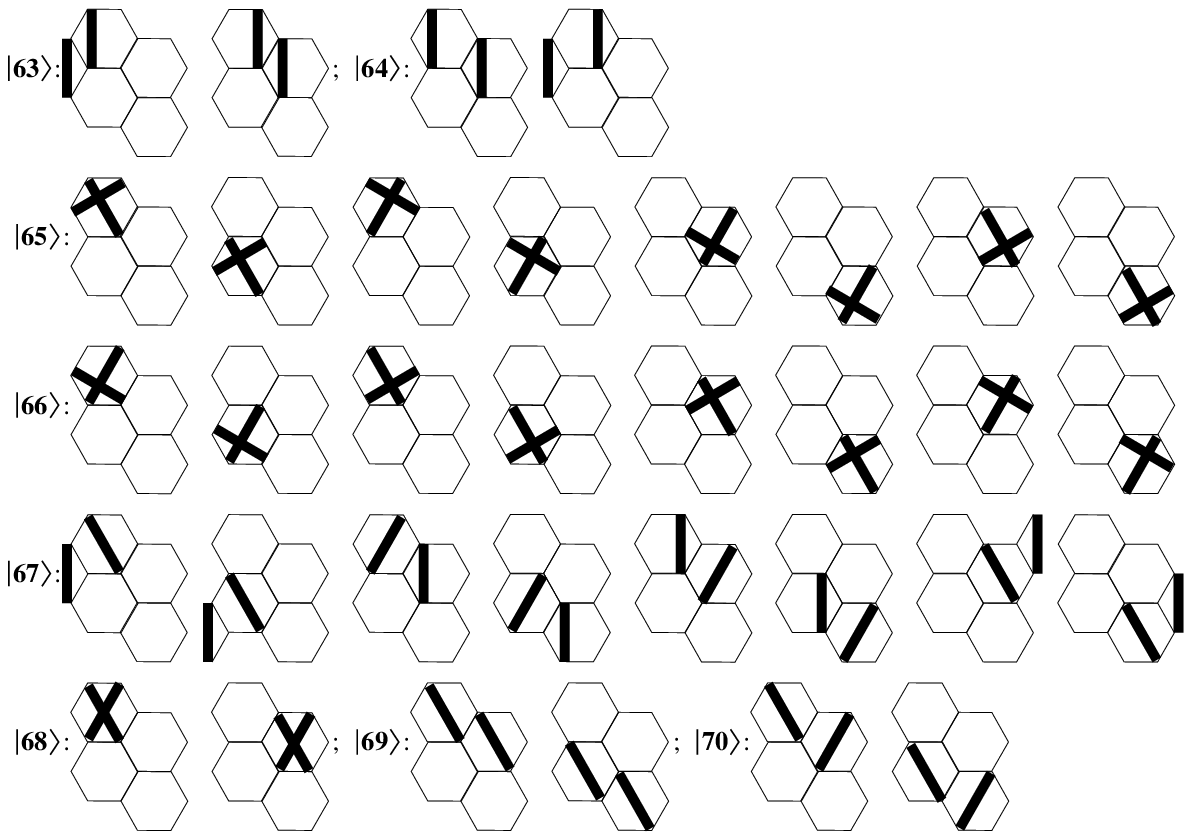}} %
\caption{The structure of the base vectors $|63\rangle-|70\rangle$ of    %
${\cal{S}}$. The meaning of the notations is as in Figs.(3-5).}          %
\end{figure}                                                             %

\newpage

In order to reproduce the ground state, from Eq.(A1) one expresses the
eigenvector providing the smallest energy. The fact that we indeed find the
ground state from Appendix A, has been checked by the exact diagonalization in 
the 784 dimensional full Hilbert space. The obtained ground state energy 
values, $E_g$ (which are the same in both ${\cal{H}}$ and ${\cal{S}}$) are 
presented below for different $\hat H$ parameters in Tables 1-3, where 
all quantities are expressed in $t$ units.

\vspace*{0.5 cm}
\begin{center}
\begin{tabular}{||c c||}
\hline
\hline
\multicolumn{2}{||c||}{$t'=0$}\\\hline
\multicolumn{1}{||c}{}$U$ \vline & $E_g$\\\hline
0.0 \vline & -8.000000000000 \\
0.5 \vline & -7.826052697604 \\
1.0 \vline & -7.675901871093 \\
1.5 \vline & -7.545391958586 \\
2.0 \vline & -7.431230836069 \\
2.5 \vline & -7.330781976775 \\
3.0 \vline & -7.241912968838 \\
3.5 \vline & -7.162884307355 \\
4.0 \vline & -7.092266429238 \\
4.5 \vline & -7.028876746763 \\
5.0 \vline & -6.971731130272 \\\hline\hline
\end{tabular}
\hspace*{0.5cm}
\begin{tabular}{||c c||}
\hline
\hline
\multicolumn{2}{||c||}{$t'=0.1$}\\\hline
\multicolumn{1}{||c}{}$U$ \vline & $E_g$\\\hline
0.0 \vline & -7.200000000000 \\
0.5 \vline & -7.029096523521 \\
1.0 \vline & -6.886663391590 \\
1.5 \vline & -6.766883213589 \\
2.0 \vline & -6.665278322743 \\
2.5 \vline & -6.578374280956 \\
3.0 \vline & -6.503455227928 \\
3.5 \vline & -6.438383639055 \\
4.0 \vline & -6.381465768299 \\
4.5 \vline & -6.331350302916 \\
5.0 \vline & -6.286951600765 \\\hline\hline
\end{tabular}
\hspace*{0.5cm}
\begin{tabular}{||c c||}
\hline
\hline
\multicolumn{2}{||c||}{$t'=0.5$}\\\hline
\multicolumn{1}{||c}{}$U$ \vline & $E_g$\\\hline
0.0 \vline & -8.000000000000 \\
0.5 \vline & -7.826554868506 \\
1.0 \vline & -7.678117036240 \\
1.5 \vline & -7.550564402476 \\
2.0 \vline & -7.440430706187 \\
2.5 \vline & -7.344831780650 \\
3.0 \vline & -7.261386257598 \\
3.5 \vline & -7.188137025326 \\
4.0 \vline & -7.123478675601 \\
4.5 \vline & -7.066093843884 \\
5.0 \vline & -7.014899323332 \\\hline\hline
\end{tabular}
\\
\vspace*{1cm}
Table 1.
\hspace{3cm}
Table 2.
\hspace{3cm}
Table 3.
\end{center}

\subsection{Observations relating to the applied procedure}

From Eq.(\ref{Eq7}) it can be observed that the starting vector $|1\rangle$,
by the action of the Hamiltonian, reproduces also the vectors $|2\rangle$,
$|3\rangle$, which are similar to $|1\rangle$ and can be considered also as of
``most interacting configuration'' type. Indeed, the whole Eq.(A1) system of 
equations can be reproduced starting from the vector $|2\rangle$ or vector 
$|3\rangle$. Note that the impression that these last two vectors have 
non-parallel (i.e. rotated) contributions is misleading. Indeed, for both 
$|2\rangle$ and $|3\rangle$, the first two contributions are placed on the outer
circumference of the torus, while the second two contributions on the inner
circumference of the torus. Consequently, both vectors $|2\rangle$ and 
$|3\rangle$ are built up only from contributions which are parallel inside the 
sample.

The study of Figs.(3-8) shows that several possible particle configurations are
missing from the ground state (similar property holds also for the square 
system, see Ref.\cite{I1}). This is because only those configurations are
present in $|\Psi_g\rangle$ which, by the action of the Hamiltonian, can be 
connected to the most interacting configuration. That is why, in constructing
the base vectors of ${\cal{S}}$ (i.e. Appendix A), we must use a starting 
vector which describes the most interacting configuration. 

We note that in case of the square system described in Ref.\cite{I1},
all vectors describing a most-interacting configuration (around a given lattice
site, all these can be obtained from each other by a rotation with $\pi/2$), 
appear always with the same coefficient, so can be added together in a unique 
starting vector. In our case, however, the vectors $|1\rangle$, $|2\rangle$, 
$|3\rangle$
do not have this property (i.e. vectors $|1\rangle$, $|2\rangle$, $|3\rangle$ 
are 
separated and can not be added together), because the sample we use, does not 
possess $2\pi/3$
rotational symmetry. Practically this is the reason why one reaches in the 
studied
case only one order of magnitude decrease in reducing $Dim({\cal{H}})$ to
$Dim({\cal{S}})$ in the process of deducing the ground state.
For clarity, the detailed construction of the components of an arbitrary
vector $|i\rangle$ at fixed $i$ in the studied case is presented in Appendix B.

\section{Properties of the deduced ground state}

By studying the deduced properties, first one notes that the system in 
Appendix A properly reproduces the ground state, but it is incomplete at the
level of excited states. Since several low lying excited states are not
provided by Eq.(\ref{A1}), the reduced space ${\cal{S}}$ cannot be used for
the study of excitations, or for the estimation of the charge gap.

Turning back to the ground state, with its
explicit expression deduced from the reduced
${\cal{S}}$ subspace, several ground state properties of the 
system can be analyzed. 

One often finds continuously increasing singlet ground state energy for 
Hubbard models at increasing $U$  on finite $U$ domains in one
\cite{I2,I2a,I2b} and two \cite{I3} dimensions as well, and even if we know that
in 1D, the Bethe Ansatz $E_g$ result saturates at $U \to \infty$, see Ref.
\cite{I4}, we also know that often, the emergence of ferromagnetism at a fixed
concentration is associated with the singlet $E_g$ increase in function of 
increasing $U$ \cite{I4a}.
 
Consequently, taken into account that the most interacting configuration 
(i.e. the configuration containing the maximum number of double occupancies
dense displaced) enters our ground state, one naively expects that 
if $U$ increases, the 
singlet four-particle ground state energy also continuously increases. 
The result however shows that $E_g$ reaches a saturation when $U$ increases 
(see Fig.9.a), and the system remains in singlet state even at 
$U \to \infty$. For the 2D case, in exact terms, a such kind of saturation 
in function of the
interaction, in our knowledge, has not been shown yet. The presented
property is not connected exclusively to graphene-like systems, since it 
appears also for 2D square lattice (see Fig.9.b). This last figure has been 
deduced based on the results \cite{I5} published in Ref.\cite{I1}.

\begin{figure} [h]                                                   %
\centerline{\includegraphics[width=16cm,height=6cm]{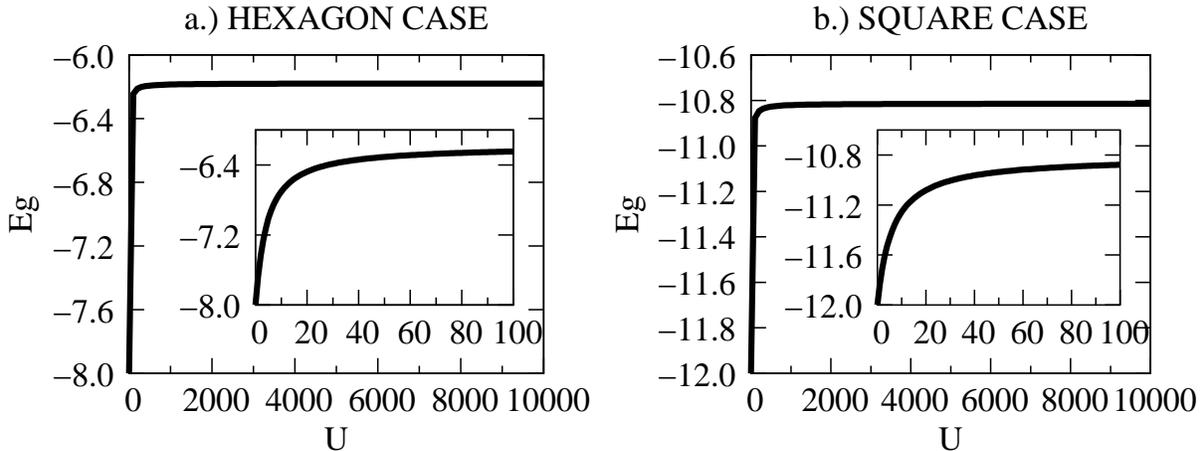}} %
\caption{The ground state energy in function of $U$. a) With hexagonal 
repeat units, the case presented in Fig.1 at $t'/t=0.5$. b) 4x4 square 
lattice studied in Ref.\cite{I1}, where $U$ is in $t$ units.} 
\end{figure}                                                         %

It turned out that the observed saturation emerges because, even if the most 
interacting particle configurations (i.e. $|1\rangle, |2\rangle, |3\rangle$) 
are present in the normalized ground state wave function
\begin{eqnarray}
|\Psi_g\rangle= \sum_{i=1}^{70} x_i |i\rangle,
\label{Eq8}
\end{eqnarray}
the coefficients $x_i$ of the basis vectors containing double occupancies
strongly decrease with increasing $U$. Indeed, Fig.10 shows the dependence 
on $U$ of
the $|1\rangle$ base vector containing two nearest neighbor double 
occupancies in a non-degenerate ground state provided by $\hat H$ in 
(\ref{Eq1}).

\begin{figure} [h]                                                   %
\centerline{\includegraphics[width=8cm,height=6.4cm]{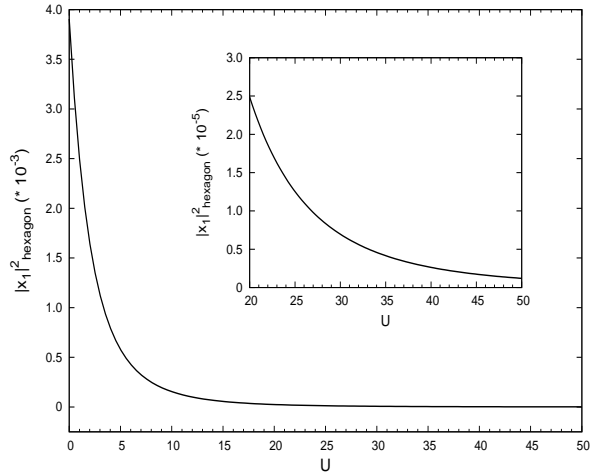}} %
\caption{The dependence of the $|x_1|^2$ coefficient of (\ref{Eq8}) in the
ground state of the system with hexagonal repeat units in function of $U$. 
The ground state is non-degenerate and corresponds to $t'/t=-0.6$.
The U values are expressed in t units. }
\end{figure}                                                         %

\begin{figure} [h]                                                   %
\centerline{\includegraphics[width=8cm,height=6.4cm]{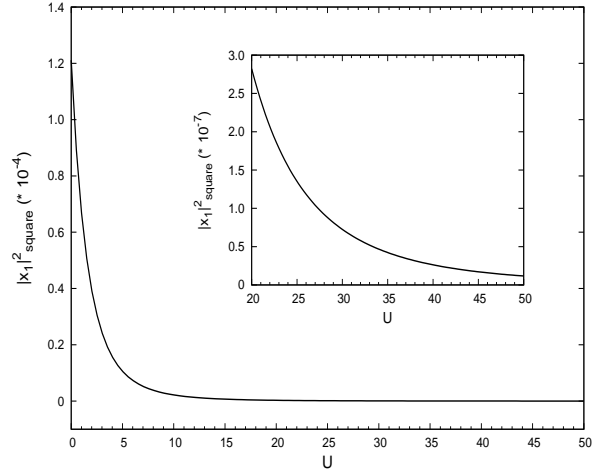}} %
\caption{The dependence on $U$ of the $|x_{1}|^2$ value in case of the 
square system studied in Ref.\cite{I1} (the notation of $|1\rangle$ is taken
from Ref.\cite{I1}). The corresponding vector 
contains two double occupancies on nearest neighbor sites.
$U$ is expressed in $t$ units.} 
\end{figure}                                                         %

As seen, the decrease is strong, and one finds similar behavior also in the
square system, see Fig.11. Compairing Figs.10-11, one sees that the
emergence probabilities of configurations with two double occupancies in
2D systems with hexagonal and square repeat units, in the presented case,
behave similar, and their decrease rate in function of U is also similar
\cite{expl1}.

Up to this moment the behavior and effects of the interaction in systems with
hexagonal and square repeat units seem to be resembling. However, what makes a 
system with hexagonal repeat units different from the square one, is the 
emergence of closely placed low lying energy levels which lead to degenerate
(or almost degenerate) ground states in extended regions of the phase diagram.
This is observed also in other studies relating honeycomb systems
\cite{I4x,I4y,I4z}. This situation will be exemplified below (see Figs.12-13)
for a ground state $|\Psi_g\rangle = |\Psi_{g,1}\rangle$, whose energy 
$E_g=E_{g,1} \leq E_{g,2}$, within the numerical error of the calculation, 
coincides to the energy $E_{g,2}$ provided by the nearest neighbor level
described by $|\Psi_{g,2}\rangle$. Note that for $n=1,2$, the vectors
$|\Psi_{g,n}\rangle$ are ortho-normalized. In this case,
the emergence probability of different particle configurations in
$|\Psi_g\rangle$ shows trembling in function of U. For exemplification 
we present for the start two plots, namely
first the $U$ dependence of the $|x_1|^2$ coefficient in Fig.12, and
second, the $U$ dependence of $|x_8|^2$ coefficient in Fig.13.

\begin{figure} [h]                                                   %
\centerline{\includegraphics[width=8cm,height=6.4cm]{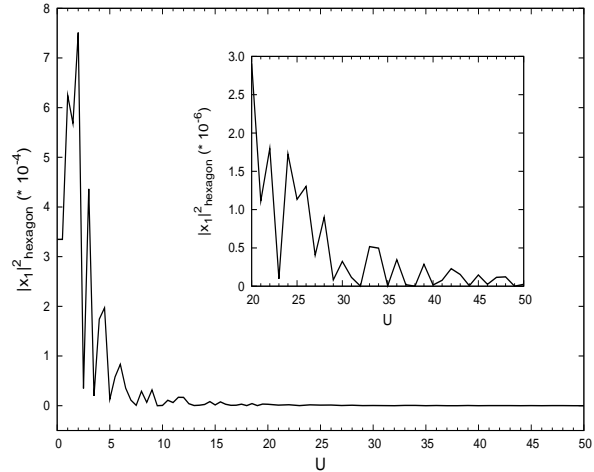}} %
\caption{The dependence on $U$ of the $|x_1|^2$ value from
Eq.(\ref{Eq8}) in the case of the system with hexagonal repeat 
units presented in Fig.1. The corresponding vector 
contains two double occupancies on nearest neighbor sites. One has 
$t'/t=0.5$, and $U$ is expressed in $t$ units.} 
\end{figure}                                                         %

\begin{figure} [h]                                                   %
\centerline{\includegraphics[width=8cm,height=6.4cm]{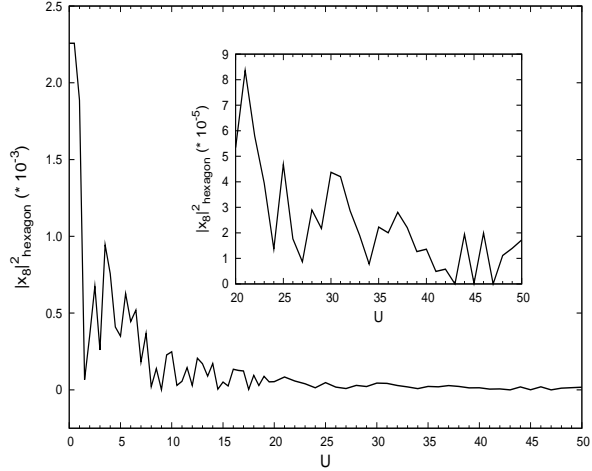}} %
\caption{The dependence on $U$ of the $|x_8|^2$ value 
from Eq.(\ref{Eq8}) in the case of the system 
with hexagonal repeat units presented in Fig.1. The corresponding vector 
contains a double occupancy and two single occupancies on nearest neighbor 
sites in nearest neighbor position from the double occupied site. 
One has $t'/t=0.5$, and $U$ is expressed in $t$ units.} 
\end{figure}                                                         %

One notes that the basis vector $|1\rangle$ corresponding to the $|x_1|^2$
coefficient (see Fig.12) contains two double occupancies placed in 
neighboring positions, while the basis vector $|8\rangle$ connected to
the $|x_8|^2$ coefficient (see Fig.13) contains only one double occupancy 
and two single occupancies on nearest neighbor sites in nearest neighbor 
position from the double occupied site (see Fig.3). 
In case of Fig.12, the shape of the function at 
$U \to \infty$ is similar to Fig.13, but now
a maximum is reached at $U=2$. The presence of a clear trembling 
in the $U$ dependence is clearly seen in both cases.
For the same conditions, 
similar behavior is seen in other $|x_i|^2$ coefficients 
relating $|i\rangle$ states contained in $|\Psi_{g}\rangle$.
In order to exemplify, we present in Figs.14-15 two more cases, the
first being related to two double occupancies placed on next nearest
neighbor positions (Fig.14), and the second being connected to one double
occupancy and two single occupancies on nearest neighbor 
sites placed in next nearest neighbor
position from the double occupied site (Fig.15).

\begin{figure} [h]                                                   %
\centerline{\includegraphics[width=8cm,height=6.4cm]{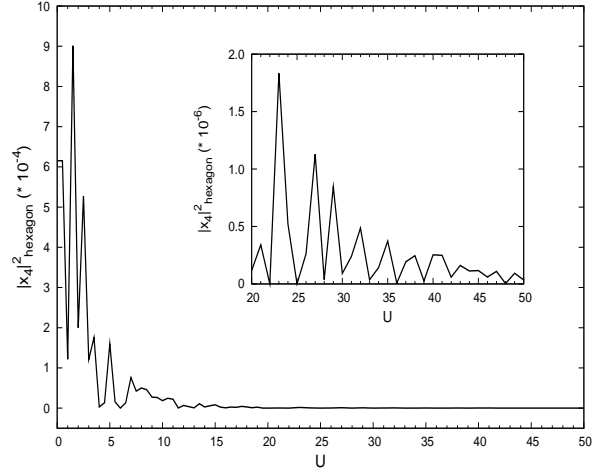}} %
\caption{The dependence on $U$ of the $|x_4|^2$ value from Eq.(\ref{Eq8})
in the case of the system 
with hexagonal repeat units presented in Fig.1. The corresponding vector 
contains two double occupancies placed on next nearest neighbor 
sites. One has $t'/t=0.5$, and $U$ is expressed in $t$ units.} 
\end{figure}                                                         %

\begin{figure} [h]                                                   %
\centerline{\includegraphics[width=8cm,height=6.4cm]{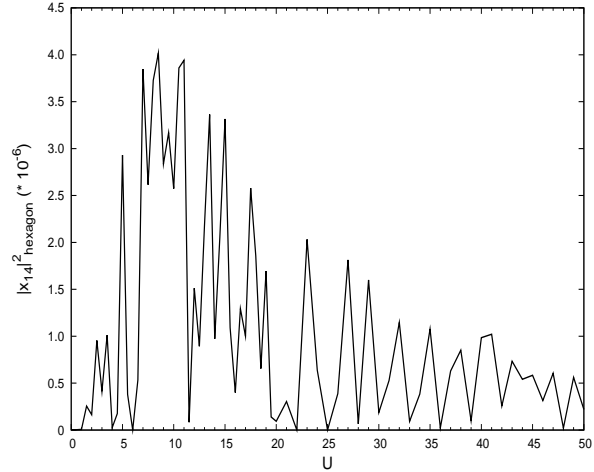}} %
\caption{The dependence on $U$ of the $|x_{14}|^2$ value 
from Eq.(\ref{Eq8}) in the case of the system 
with hexagonal repeat units presented in Fig.1. The corresponding vector 
contains one double occupancy and two single occupancies on nearest neighbor 
sites in next nearest neighbor position from the double occupied site.
One has $t'/t=0.5$, and $U$ is expressed in $t$ units.} 
\end{figure}                                                         %

As seen,
if the distances between two double occupancies or between a double and a 
pair of single occupancies in the particle configurations are increased, see 
Figs.14-15, the trembling character of the behavior 
and the decrease in function of $U$ at $U \to \infty$ remains, but
the value of $|x_i|^2$ is in the same time strongly 
diminishes. Similarly to Fig.12, a maximum value can be observed in Fig.14 at 
$U=1.5$, and in Fig.15 at $U=8.5$. We note that trembling has been observed 
also at $t'=0$.

In order to underline that the trembling is missing in the square case, one
presents below three examples in Figs.16-18, namely the case of a double
occupancy and two single occupancies on nearest neighbor sites in 
nearest neighbor position from the double occupied site (Fig.16),
the case of two double occupancies placed in third neighbor positions
(Fig.17), and finally, the case of one
double occupancy and two single occupancies on nearest 
neighbor sites placed in third neighbor 
position from the double occupied site (Fig.18).

\begin{figure} [h]                                                   %
\centerline{\includegraphics[width=8cm,height=6.4cm]{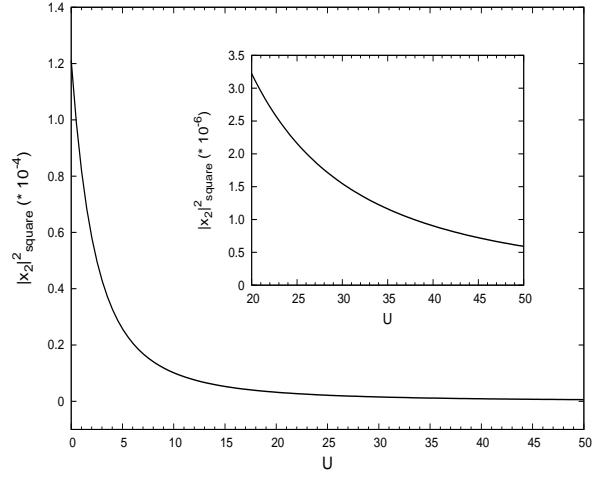}} %
\caption{The dependence on $U$ of the $|x_{2}|^2$ value in case of the 
square system studied in Ref.\cite{I1} (the notation of $|2\rangle$ is
taken from Ref.\cite{I1}). The corresponding vector 
contains one double occupancy and two single occupancies on nearest 
neighbor sites in nearest neighbor position from the double occupied site. 
$U$ is expressed in $t$ units.} 
\end{figure}                                                         %

\begin{figure} [h]                                                   %
\centerline{\includegraphics[width=8cm,height=6.4cm]{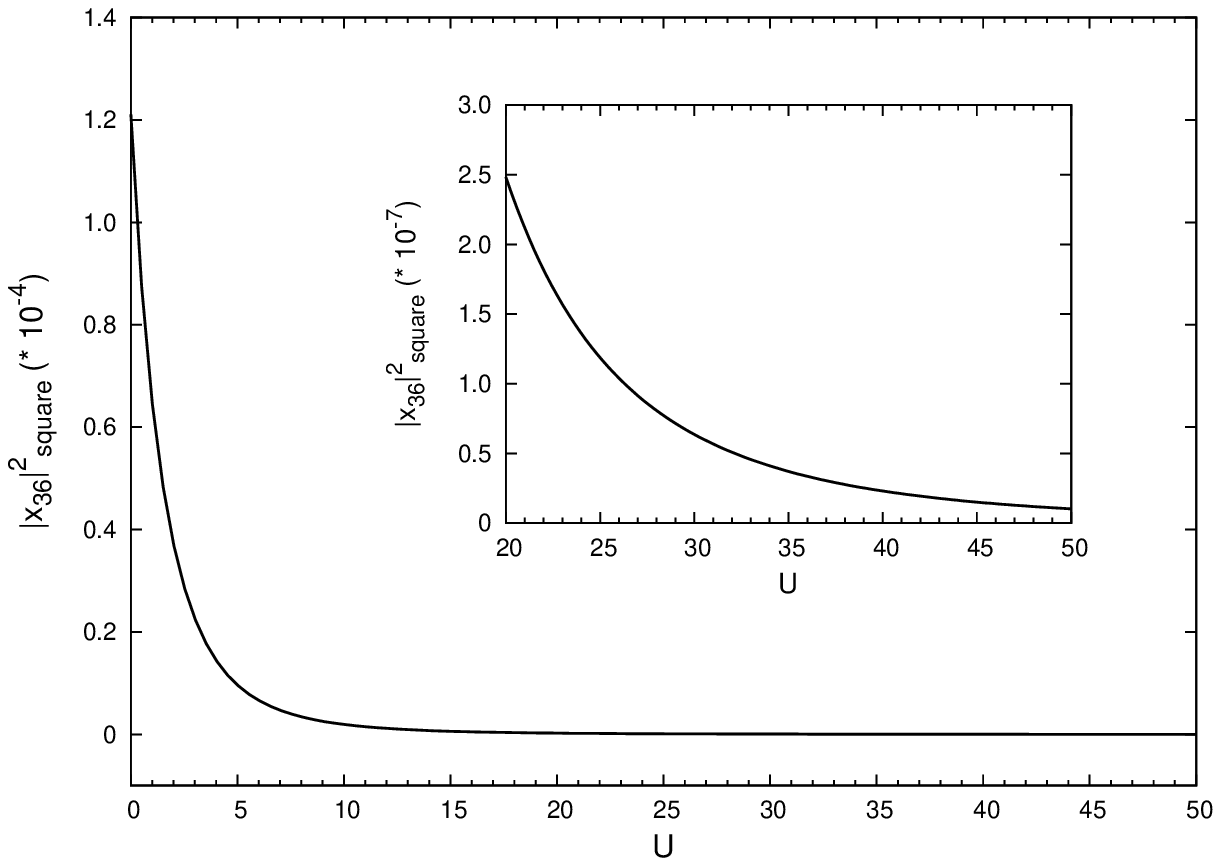}} %
\caption{The dependence on $U$ of the $|x_{36}|^2$ value in case of the 
square system studied in Ref.\cite{I1} (the notation of $|36\rangle$ is 
taken from Ref.\cite{I1}). The corresponding vector 
contains two double occupancies on third neighbor sites.
$U$ is expressed in $t$ units.} 
\end{figure}                                                         %

\begin{figure} [h]                                                   %
\centerline{\includegraphics[width=8cm,height=6.4cm]{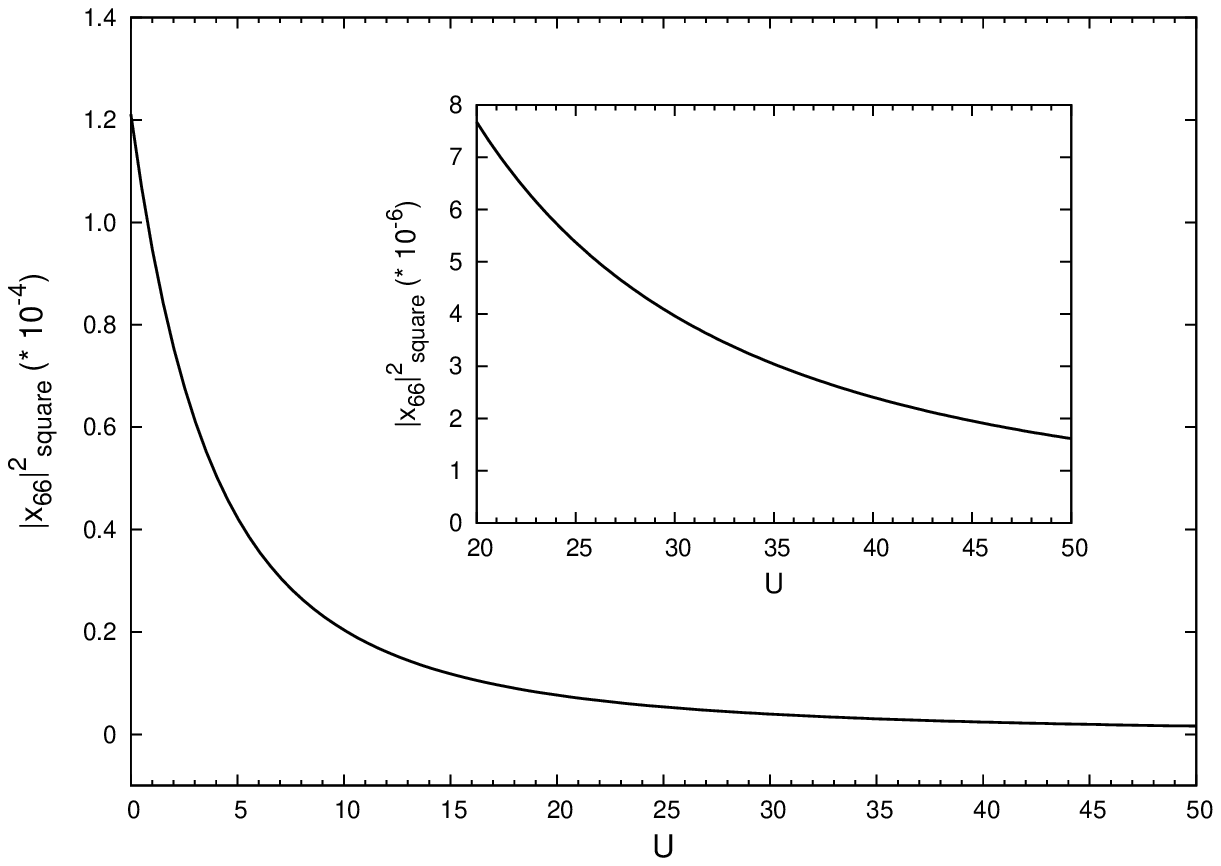}} %
\caption{The dependence on $U$ of the $|x_{66}|^2$ value in case of the 
square system studied in Ref.\cite{I1} (the notation of $|66\rangle$ is 
taken from Ref.\cite{I1}). The corresponding vector 
contains one double occupancy and two single occupancies on nearest 
neighbor sites in third neighbor position from the double occupied site.
$U$ is expressed in $t$ units.} 
\end{figure}                                                         %

\begin{figure} [h]                                                   %
\centerline{\includegraphics[width=8cm,height=6.4cm]{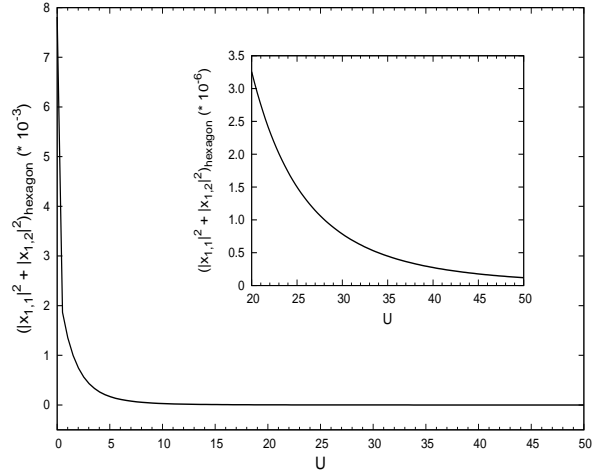}} %
\caption{The $U$ dependence of the $\sum_{n=1}^2|x_{1,n}|^2=
|x_{1,1}|^2+|x_{1,2}|^2$ sum in the hexagon case, $i=1$ particle configuration
and $t'/t=0.5$ double degenerate situation. As seen, the trembling in
$U$ is missing. The $U$ value is expressed in $t$ units.} 
\end{figure}                                                         %

Comparing the results deduced for hexagonal repeat units with the case 
of the square lattice (see Figs.16-18), one observes that
the decrease of the $|x_i|^2$ coefficients in function of $U$ remains,
but the trembling is missing, and the maximum disappears.

The trembling (``Zittern'' in German language) is known mostly because
of the trembling motion (``Zitterbewegung'') of the Dirac electron, namely
the trembling of the relativistic electron velocity (hence also the
electron position) in function of time \cite{Z1} observed by Schr\"odinger
(see for the original derivation Ref.\cite{Z2}). However, it is clear that
trembling is not connected to relativity, since can occur also in classical 
wave propagation phenomena \cite{Z3}. In the present case the trembling 
occurs not in a time dependent phenomenon, but in the $U$ dependence of the
emergence probability $|x_i|^2$ of a particle configuration $i$ described by
the state vector $|i\rangle$ present in the ground state. 

The trembling appears (as in the Dirac electron case) because of an 
interference between two states influencing each other in the frame of the
concrete event (particle and antiparticle states in Zitterbewegung of Dirac 
electrons). In the present case the interference is caused by the 
proximity of two states on the energy scale. In order to check this statement,
if one calculates the $\sum_{n=1,2}|x_{1,n}|^2$ quantity taking $x_{i,n}$ from 
$|\Psi_{g,n}\rangle$, $n=1,2$, one finds a continuous (i.e. trembling free)
behavior, as observed from Fig.19.

As seen from Figs.16-19, for relatively small U values,
an oscillatory contribution is present in trembling (such behavior is present
also in the relativistic Zitterbewegung), whose period is close to
the $t$ value (note that U is measured in t units), but this is transient, 
since disappears in $U \to \infty$ limit (see for example the high U region of
Fig.14).

We note that if $|\Psi_{g,n}\rangle$, $n=1,2$ describe a rigorously degenerate
state, the described trembling behavior remains present in a realistic system.
This is because even under the action of an infinitesimally small perturbation, 
the degeneracy is broken (see for example the case of a short ranged impurity 
\cite{I4x}). Indeed, let us consider $x_{i,n}$, $n=1,2$, the coefficients of 
the particle configuration $i$ described by the state vector $|i\rangle$ in 
the degenerate ground state $|\Psi_{g,n}\rangle$. One knows that $x_{i,n}$ are 
trembling, but as observed from Fig.19, the function $f(U)$ defined by
\begin{eqnarray}
|x_{i,1}|^2+|x_{i,2}|^2=f(U)
\label{Eq9}
\end{eqnarray}
is a continuous non-trembling (i.e. possessing
continuous U derivative) function. Then, from the stationary degenerate 
perturbation theory one knows that the emerging non-degenerate ground state
$|\Psi_g\rangle$ becomes a linear combination of $|\Psi_{g,n}\rangle$ vectors 
with fixed prefactors. Hence in $|\Psi_g\rangle$, the vector $|i\rangle$
has the coefficient $x_i= x_{i,1} + K x_{i,2}$, where K is fixed, and is 
explicitly determined by the degenerate perturbation theory. It depends 
in fact on the matrix elements of the perturbation expressed in terms of 
the non-perturbed eigenstates belonging to the
degenerate level. It can be seen that because of (\ref{Eq9}), in 
$|x_i|^2=|x_{i,1} + K x_{i,2}|^2$
the trembling will be automatically preserved.

The deduced results show that, contrary to square lattices, in 2D systems with 
hexagonal repeat units strong variations in the system are possible to appear
following small, even infinitesimal modifications in the value of the 
interaction. Given by this property, the interaction dependent behavior of a
honeycomb system as graphene could substantially differ from the behavior of a
square lattice, even for concentrations which place far away the Fermi level
from the Dirac points. Such in principle differences in the behavior could  
cause controversies as encountered in Refs.[\cite{Intr18,Intr19,Intr20}].

One notes that the presented technique can be applied also in the presence of
non-local interactions. On this line we expect that density-density
type of non-local interactions essentially will not modify the observed 
properties. Furthermore, in the presence of local interactions, the brick-wall
lattice (see for example Ref.[\cite{Z4}]) at $t'=0$ has the spectrum 
of the honeycomb lattice. Taking next-nearest neighbor hoppings into account, 
differences appear relative to the honeycomb case, because $t'_1, t'_2$ must 
be defined instead of a single next-nearest neighbor $t'$ hopping term. 
However, we do not expect that this difference will alter in main aspects the 
behavior described in this paper.

\section{Summary and discussions}

We deduced the exact interacting four particle ground state of a 2D finite 
sample described by a Hubbard type of model and build up from hexagon repeat 
units. The ground state is obtained exactly 
from a restricted space ${\cal{S}}$ with dimensionality much smaller than the
dimension of the full Hilbert space ${\cal{H}}$ of the problem,
$D_{\cal{S}}=Dim({\cal{S}}) << Dim({\cal{H}})=D_{\cal{H}}$. 
The used technique begins
from a starting wave vector $|1\rangle$ containing the most interacting 
particle configuration (i.e. two nearest neighbor double occupancies) 
translated to all sublattice sites and added). The application of the 
Hamiltonian ($\hat H$) on $|1\rangle$ leads to further vectors $|i\rangle$
with similar properties, i.e. a local particle configuration translated to
different sites and added. Taken together, the 
$\hat H |i\rangle=\sum_j \alpha_{j,i} |j\rangle$ equalities build up a
closed system of linear equations for $i \leq D_{\cal{S}}<<D_{\cal{H}}$, whose
minimum energy solution represents the ground state. We note that the ground
state was always found a non-magnetic singlet state.

With the exact ground state in hands, different properties of the system
have been analyzed. We have found that contrary to expectations, the singlet
ground state energy saturates in the limit of infinite on-site Coulomb 
repulsion, and the emergence probability of different particle configurations
in the ground state presents trembling (``Zittern'') in function of $U$, this
property being absent in the case of a square lattice. The trembling behavior
has been shown to appear because of the interference between states placed in 
the proximity of each other on the energy scale. It can lead to strong 
modifications of the system properties caused by small  variations 
of the interaction strength, and can be the source of the differences in     
the interaction dependent behavior of square and honeycomb 2D systems.

\section{Acknowledgments}

Z.G. kindly acknowledges financial support provided by Alexander von 
Humboldt Foundation, OTKA-K-100288 (Hungarian Research Funds for Basic 
Research), and TAMOP 4.2.2/A-11/1/KONV-2012-0036 (co-financed by EU 
and European Social Fund).

\appendix

\section{The system of equations providing the ground state in 
the 70 dimensional subspace ${\cal{S}}$.}

\begin{eqnarray}
\hat H |1\rangle &=& 2U |1\rangle + 2t |8\rangle + 2t |10\rangle + 4t' 
|17\rangle + 4t' |18\rangle + 4t' |22\rangle,
\nonumber\\
\hat H |2\rangle &=& 2U |2\rangle + 2t |11\rangle + 2t |13\rangle + 4t' 
|18\rangle + 4t' |19\rangle + 4t' |21\rangle,
\nonumber\\
\hat H |3\rangle &=& 2U |3\rangle + 2t |9\rangle + 2t |12\rangle + 4t' 
|17\rangle + 4t' |19\rangle + 4t' |20\rangle,
\nonumber\\
\hat H |4\rangle &=& 2U |4\rangle + 2t |8\rangle + 2t |12\rangle + 2t |14\rangle + 4t' |23\rangle + 4t' |25\rangle,
\nonumber\\
\hat H |5\rangle &=& 2U |5\rangle + 2t |10\rangle + 2t |13\rangle + 2t |15\rangle + 4t' |24\rangle + 4t' |25\rangle,
\nonumber\\
\hat H |6\rangle &=& 2U |6\rangle + 2t |9\rangle + 2t |11\rangle + 2t |16\rangle + 4t' |23\rangle + 4t' |24\rangle,
\nonumber\\
\hat H |7\rangle &=& 2U |7\rangle + 2t |14\rangle + 2t |15\rangle + 2t |16\rangle + 4t' |20\rangle + 4t' |21\rangle + 4t' |22\rangle,
\nonumber\\
\hat H |8\rangle &=& 2t |1\rangle + 2t |4\rangle + U |8\rangle + 2t' |10\rangle + 2t' |14\rangle + t |22\rangle + t |25\rangle + 
2t' |27\rangle + 2t' |28\rangle 
\nonumber\\
&-& 2t |37\rangle - 2t' |42\rangle - 2t' |43\rangle - 2t' |47\rangle + t |52\rangle - 2t |53\rangle + t |57\rangle - 2t' |65\rangle,
\nonumber\\
\hat H |9\rangle &=& 2t |3\rangle + 2t |6\rangle + U |9\rangle + 2t' |12\rangle + 2t' |16\rangle + t |20\rangle + t |23\rangle + 
2t' |26\rangle + 2t' |28\rangle 
\nonumber\\
&+& 2t |35\rangle + 2t' |41\rangle + 2t' |47\rangle - 2t' |48\rangle + t |50\rangle + t |55\rangle + 2t |63\rangle + 2t' |67\rangle,
\nonumber\\
\hat H |10\rangle &=& 2t |1\rangle + 2t |5\rangle + 2t' |8\rangle + U |10\rangle + 2t' |15\rangle + t |22\rangle + t |25\rangle + 
2t' |26\rangle + 2t' |28\rangle 
\nonumber\\
&-& 2t |38\rangle + 2t' |41\rangle - 2t' |44\rangle + 2t' |47\rangle + t |52\rangle + 2t |54\rangle + t |57\rangle + 2t' |66\rangle,
\nonumber\\
\hat H |11\rangle &=& 2t |2\rangle + 2t |6\rangle + U |11\rangle + 2t' |13\rangle + 2t' |16\rangle + t |21\rangle + t |24\rangle + 
2t' |27\rangle + 2t' |28\rangle 
\nonumber\\
&-& 2t |36\rangle - 2t' |42\rangle - 2t' |47\rangle - 2t' |49\rangle - t |51\rangle - t |56\rangle - 2t |63\rangle - 2t' |67\rangle,
\nonumber\\
\hat H |12\rangle &=& 2t |3\rangle + 2t |4\rangle + 2t' |9\rangle + U |12\rangle + 2t' |14\rangle + t |20\rangle + t |23\rangle + 
2t' |26\rangle + 2t' |27\rangle 
\nonumber\\
&+& 2t |29\rangle - 2t' |41\rangle + 2t' |42\rangle + 2t' |45\rangle + t |50\rangle + 2t |53\rangle + t |55\rangle + 2t' |65\rangle,
\nonumber\\
\hat H |13\rangle &=& 2t |2\rangle + 2t |5\rangle + 2t' |11\rangle + U |13\rangle + 2t' |15\rangle + t |21\rangle + t |24\rangle + 
2t' |26\rangle + 2t' |27\rangle 
\nonumber\\
&-& 2t |30\rangle - 2t' |41\rangle + 2t' |42\rangle - 2t' |46\rangle - t |51\rangle - 2t |54\rangle - t |56\rangle - 2t' |66\rangle,
\nonumber\\
\hat H |14\rangle &=& 2t |4\rangle + 2t |7\rangle + 2t' |8\rangle + 2t' |12\rangle + U |14\rangle + 2t' |15\rangle + 2t' |16\rangle + t |20\rangle + t |22\rangle 
\nonumber\\
&+& t |23\rangle + t |25\rangle - 2t' |44\rangle - 2t' |48\rangle + t |50\rangle + t |52\rangle + t |55\rangle + t |57\rangle - 
2t |61\rangle 
\nonumber\\
&+& 2t' |66\rangle + 2t' |67\rangle + 2t |69\rangle,
\nonumber\\
\hat H |15\rangle &=& 2t |5\rangle + 2t |7\rangle + 2t' |10\rangle + 2t' |13\rangle + 2t' |14\rangle + U |15\rangle + 2t' |16\rangle + t |21\rangle + t |22\rangle 
\nonumber\\
&+& t |24\rangle + t |25\rangle - 2t' |43\rangle - 2t' |49\rangle - t |51\rangle + t |52\rangle - t |56\rangle + t |57\rangle - 
2t |62\rangle 
\nonumber\\
&-& 2t' |65\rangle - 2t' |67\rangle - 2t |70\rangle,
\nonumber
\end{eqnarray}

\newpage

\begin{eqnarray}
\hat H |16\rangle &=& 2t |6\rangle + 2t |7\rangle + 2t' |9\rangle + 2t' |11\rangle + 2t' |14\rangle + 2t' |15\rangle + U |16\rangle + t |20\rangle + t |21\rangle 
\nonumber\\
&+& t |23\rangle + t |24\rangle + 2t' |45\rangle - 2t' |46\rangle + t |50\rangle - t |51\rangle + t |55\rangle - t |56\rangle - 
2t |58\rangle 
\nonumber\\
&+& 2t' |65\rangle - 2t' |66\rangle - 2t |68\rangle,
\nonumber\\
\hat H |17\rangle &=& 4t' |1\rangle + 4t' |3\rangle + U |17\rangle + 2t' |18\rangle + 2t' |19\rangle + 2t' |20\rangle + 
2t' |22\rangle + t |26\rangle + t |28\rangle 
\nonumber\\
&-& 4t' |29\rangle + 2t' |32\rangle + 2t' |34\rangle + 4t' |37\rangle + t |41\rangle + t |47\rangle - 2t' |50\rangle - 2t' |52\rangle,
\nonumber\\
\hat H |18\rangle &=& 4t' |1\rangle + 4t' |2\rangle + 2t' |17\rangle + U |18\rangle + 2t' |19\rangle + 2t' |21\rangle + 
2t' |22\rangle + t |27\rangle + t |28\rangle 
\nonumber\\
&+& 4t' |30\rangle + 2t' |31\rangle + 2t' |34\rangle + 4t' |38\rangle - t |42\rangle - t |47\rangle + 2t' |51\rangle - 2t' |52\rangle,
\nonumber\\
\hat H |19\rangle &=& 4t' |2\rangle + 4t' |3\rangle + 2t' |17\rangle + 2t' |18\rangle + U |19\rangle + 2t' |20\rangle + 
2t' |21\rangle + t |26\rangle + t |27\rangle 
\nonumber\\
&+& 2t' |31\rangle + 2t' |32\rangle - 4t' |35\rangle + 4t' |36\rangle - t |41\rangle + t |42\rangle - 2t' |50\rangle + 2t' |51\rangle,
\nonumber\\
\hat H |20\rangle &=& 4t' |3\rangle + 4t' |7\rangle + t |9\rangle + t |12\rangle + t |14\rangle + t |16\rangle + 2t' |17\rangle + 
2t' |19\rangle + U |20\rangle 
\nonumber\\
&+& 2t' |21\rangle + 2t' |22\rangle + 2t' |31\rangle + 2t' |34\rangle - 4t' |39\rangle + t |45\rangle - t |48\rangle + 2t' |51\rangle 
\nonumber\\
&-& 2t' |52\rangle + t |65\rangle + t |67\rangle + 4t' |70\rangle,
\nonumber\\
\hat H |21\rangle &=& 4t' |2\rangle + 4t' |7\rangle + t |11\rangle + t |13\rangle + t |15\rangle + t |16\rangle + 2t' |18\rangle + 
2t' |19\rangle + 2t' |20\rangle 
\nonumber\\
&+& U |21\rangle + 2t' |22\rangle + 2t' |32\rangle + 2t' |34\rangle - 4t' |40\rangle - t |46\rangle - t |49\rangle - 2t' |50\rangle 
\nonumber\\
&-& 2t' |52\rangle - t |66\rangle - t |67\rangle - 4t' |69\rangle,
\nonumber\\
\hat H |22\rangle &=& 4t' |1\rangle + 4t' |7\rangle + t |8\rangle + t |10\rangle + t |14\rangle + t |15\rangle + 2t' |17\rangle + 
2t' |18\rangle + 2t' |20\rangle 
\nonumber\\
&+& 2t' |21\rangle + U |22\rangle + 2t' |31\rangle + 2t' |32\rangle - 4t' |33\rangle - t |43\rangle - t |44\rangle - 2t' |50\rangle 
\nonumber\\
&+& 2t' |51\rangle - t |65\rangle + t |66\rangle + 4t' |68\rangle,
\nonumber\\
\hat H |23\rangle &=& 4t' |4\rangle + 4t' |6\rangle + t |9\rangle + t |12\rangle + t |14\rangle + t |16\rangle + U |23\rangle + 
t |27\rangle + t |28\rangle 
\nonumber\\
&-& t |42\rangle + t |45\rangle - t |47\rangle - t |48\rangle + 4t' |59\rangle + 4t' |64\rangle + t |65\rangle + t |67\rangle,
\nonumber\\
\hat H |24\rangle &=& 4t' |5\rangle + 4t' |6\rangle + t |11\rangle + t |13\rangle + t |15\rangle + t |16\rangle + U |24\rangle + 
t |26\rangle + t |28\rangle 
\nonumber\\
&+& t |41\rangle - t |46\rangle + t |47\rangle - t |49\rangle - 4t' |60\rangle - 4t' |64\rangle - t |66\rangle - t |67\rangle,
\nonumber\\
\hat H |25\rangle &=& 4t' |4\rangle + 4t' |5\rangle + t |8\rangle + t |10\rangle + t |14\rangle + t |15\rangle + U |25\rangle + 
t |26\rangle + t |27\rangle 
\nonumber\\
&-& t |41\rangle + t |42\rangle - t |43\rangle - t |44\rangle - 4t' |59\rangle + 4t' |60\rangle - t |65\rangle + t |66\rangle,
\nonumber\\
\hat H |26\rangle &=& 2t' |9\rangle + 2t' |10\rangle + 2t' |12\rangle + 2t' |13\rangle + t |17\rangle + t |19\rangle + t |24\rangle + t |25\rangle + U |26\rangle 
\nonumber\\
&-& t |31\rangle - t |34\rangle + 2t' |43\rangle - 2t' |45\rangle + 2t' |48\rangle + 2t' |49\rangle + t |56\rangle - t |57\rangle,
\nonumber
\end{eqnarray}

\newpage

\begin{eqnarray}
\hat H |27\rangle &=& 2t' |8\rangle + 2t' |11\rangle + 2t' |12\rangle + 2t' |13\rangle + t |18\rangle + t |19\rangle + t |23\rangle + t |25\rangle + U |27\rangle 
\nonumber\\
&-& t |32\rangle - t |34\rangle + 2t' |44\rangle + 2t' |46\rangle + 2t' |48\rangle + 2t' |49\rangle - t |55\rangle - t |57\rangle,
\nonumber\\
\hat H |28\rangle &=& 2t' |8\rangle + 2t' |9\rangle + 2t' |10\rangle + 2t' |11\rangle + t |17\rangle + t |18\rangle + t |23\rangle + 
t |24\rangle + U |28\rangle 
\nonumber\\
&-& t |31\rangle - t |32\rangle + 2t' |43\rangle + 2t' |44\rangle - 2t' |45\rangle + 2t' |46\rangle - t |55\rangle + t |56\rangle,
\nonumber\\
\hat H |29\rangle &=& 2t |12\rangle - 4t' |17\rangle - 4t' |34\rangle + 2t |45\rangle + 4t' |50\rangle,
\nonumber\\
\hat H |30\rangle &=& - 2t |13\rangle + 4t' |18\rangle + 4t' |34\rangle + 2t |46\rangle + 4t' |51\rangle,
\nonumber\\
\hat H |31\rangle &=& 2t' |18\rangle + 2t' |19\rangle + 2t' |20\rangle + 2t' |22\rangle - t |26\rangle - t |28\rangle + 
2t' |32\rangle - 4t' |33\rangle 
\nonumber\\
&+& 2t' |34\rangle - 4t' |35\rangle + 4t' |38\rangle - 4t' |39\rangle - t |41\rangle - t |47\rangle - 2t' |50\rangle - 2t' |52\rangle,
\nonumber\\
\hat H |32\rangle &=& 2t' |17\rangle + 2t' |19\rangle + 2t' |21\rangle + 2t' |22\rangle - t |27\rangle - t |28\rangle + 
2t' |31\rangle - 4t' |33\rangle 
\nonumber\\
&+& 2t' |34\rangle + 4t' |36\rangle + 4t' |37\rangle - 4t' |40\rangle + t |42\rangle + t |47\rangle + 2t' |51\rangle - 2t' |52\rangle,
\nonumber\\
\hat H |33\rangle &=& - 4t' |22\rangle - 4t' |31\rangle - 4t' |32\rangle + 2t |43\rangle + 2t |44\rangle,
\nonumber\\
\hat H |34\rangle &=& 2t' |17\rangle + 2t' |18\rangle + 2t' |20\rangle + 2t' |21\rangle - t |26\rangle - t |27\rangle - 
4t' |29\rangle + 4t' |30\rangle 
\nonumber\\
&+& 2t' |31\rangle + 2t' |32\rangle - 4t' |39\rangle - 4t' |40\rangle + t |41\rangle - t |42\rangle - 2t' |50\rangle + 2t' |51\rangle,
\nonumber\\
\hat H |35\rangle &=& 2t |9\rangle - 4t' |19\rangle - 4t' |31\rangle - 2t |48\rangle + 4t' |50\rangle,
\nonumber\\
\hat H |36\rangle &=& - 2t |11\rangle + 4t' |19\rangle + 4t' |32\rangle + 2t |49\rangle + 4t' |51\rangle,
\nonumber\\
\hat H |37\rangle &=& - 2t |8\rangle + 4t' |17\rangle + 4t' |32\rangle + 2t |43\rangle - 4t' |52\rangle,
\nonumber\\
\hat H |38\rangle &=& - 2t |10\rangle + 4t' |18\rangle + 4t' |31\rangle + 2t |44\rangle - 4t' |52\rangle,
\nonumber\\
\hat H |39\rangle &=& - 4t' |20\rangle - 4t' |31\rangle - 4t' |34\rangle - 2t |45\rangle + 2t |48\rangle,
\nonumber\\
\hat H |40\rangle &=& - 4t' |21\rangle - 4t' |32\rangle - 4t' |34\rangle + 2t |46\rangle + 2t |49\rangle,
\nonumber\\
\hat H |41\rangle &=& 2t' |9\rangle + 2t' |10\rangle - 2t' |12\rangle - 2t' |13\rangle + t |17\rangle - t |19\rangle + t |24\rangle - t |25\rangle 
\nonumber\\
&-& t |31\rangle + t |34\rangle + 2t' |43\rangle - 2t' |45\rangle - 2t' |48\rangle - 2t' |49\rangle + t |56\rangle + t |57\rangle,
\nonumber\\
\hat H |42\rangle &=& - 2t' |8\rangle - 2t' |11\rangle + 2t' |12\rangle + 2t' |13\rangle - t |18\rangle + t |19\rangle - 
t |23\rangle + t |25\rangle 
\nonumber\\
&+& t |32\rangle - t |34\rangle - 2t' |44\rangle - 2t' |46\rangle + 2t' |48\rangle + 2t' |49\rangle + t |55\rangle - t |57\rangle,
\nonumber\\
\hat H |43\rangle &=& - 2t' |8\rangle - 2t' |15\rangle - t |22\rangle - t |25\rangle + 2t' |26\rangle + 2t' |28\rangle + 
2t |33\rangle + 2t |37\rangle 
\nonumber\\
&+& 2t' |41\rangle + 2t' |44\rangle + 2t' |47\rangle - t |52\rangle - t |57\rangle - 2t |60\rangle + 2t |62\rangle - 2t' |66\rangle,
\nonumber\\
\hat H |44\rangle &=& - 2t' |10\rangle - 2t' |14\rangle - t |22\rangle - t |25\rangle + 2t' |27\rangle + 2t' |28\rangle + 
2t |33\rangle + 2t |38\rangle 
\nonumber\\
&-& 2t' |42\rangle + 2t' |43\rangle - 2t' |47\rangle - t |52\rangle - t |57\rangle + 2t |59\rangle + 2t |61\rangle + 2t' |65\rangle,
\nonumber
\end{eqnarray}

\newpage

\begin{eqnarray}
\hat H |45\rangle &=& 2t' |12\rangle + 2t' |16\rangle + t |20\rangle + t |23\rangle - 2t' |26\rangle - 2t' |28\rangle + 
2t |29\rangle - 2t |39\rangle 
\nonumber\\
&-& 2t' |41\rangle - 2t' |47\rangle - 2t' |48\rangle + t |50\rangle + t |55\rangle - 2t |58\rangle + 2t |64\rangle + 2t' |67\rangle,
\nonumber\\
\hat H |46\rangle &=& - 2t' |13\rangle - 2t' |16\rangle - t |21\rangle - t |24\rangle + 2t' |27\rangle + 2t' |28\rangle + 
2t |30\rangle + 2t |40\rangle 
\nonumber\\
&-& 2t' |42\rangle - 2t' |47\rangle + 2t' |49\rangle + t |51\rangle + t |56\rangle + 2t |58\rangle + 2t |64\rangle + 2t' |67\rangle,
\nonumber\\
\hat H |47\rangle &=& - 2t' |8\rangle + 2t' |9\rangle + 2t' |10\rangle - 2t' |11\rangle + t |17\rangle - t |18\rangle - 
t |23\rangle + t |24\rangle 
\nonumber\\
&-& t |31\rangle + t |32\rangle + 2t' |43\rangle - 2t' |44\rangle - 2t' |45\rangle - 2t' |46\rangle + t |55\rangle + t |56\rangle,
\nonumber\\
\hat H |48\rangle &=& - 2t' |9\rangle - 2t' |14\rangle - t |20\rangle - t |23\rangle + 2t' |26\rangle + 2t' |27\rangle - 
2t |35\rangle + 2t |39\rangle 
\nonumber\\
&-& 2t' |41\rangle + 2t' |42\rangle - 2t' |45\rangle - t |50\rangle - t |55\rangle - 2t |59\rangle + 2t |61\rangle - 2t' |65\rangle,
\nonumber\\
\hat H |49\rangle &=& - 2t' |11\rangle - 2t' |15\rangle - t |21\rangle - t |24\rangle + 2t' |26\rangle + 2t' |27\rangle + 
2t |36\rangle + 2t |40\rangle 
\nonumber\\
&-& 2t' |41\rangle + 2t' |42\rangle + 2t' |46\rangle + t |51\rangle + t |56\rangle + 2t |60\rangle + 2t |62\rangle + 2t' |66\rangle,
\nonumber\\
\hat H |50\rangle &=& t |9\rangle + t |12\rangle + t |14\rangle + t |16\rangle - 2t' |17\rangle - 2t' |19\rangle - 2t' |21\rangle - 2t' |22\rangle + 4t' |29\rangle 
\nonumber\\
&-& 2t' |31\rangle - 2t' |34\rangle + 4t' |35\rangle + t |45\rangle - t |48\rangle - 2t' |51\rangle + 2t' |52\rangle + t |65\rangle + t |67\rangle
\nonumber\\
&-& 4t' |68\rangle + 4t' |69\rangle,
\nonumber\\
\hat H |51\rangle &=& - t |11\rangle - t |13\rangle - t |15\rangle - t |16\rangle + 2t' |18\rangle + 2t' |19\rangle + 2t' |20\rangle + 2t' |22\rangle + 4t' |30\rangle 
\nonumber\\
&+& 2t' |32\rangle + 2t' |34\rangle + 4t' |36\rangle + t |46\rangle + t |49\rangle - 2t' |50\rangle - 2t' |52\rangle + t |66\rangle + t |67\rangle
\nonumber\\
&+& 4t' |68\rangle + 4t' |70\rangle,
\nonumber\\
\hat H |52\rangle &=& t |8\rangle + t |10\rangle + t |14\rangle + t |15\rangle - 2t' |17\rangle - 2t' |18\rangle - 2t' |20\rangle - 2t' |21\rangle - 2t' |31\rangle 
\nonumber\\
&-& 2t' |32\rangle - 4t' |37\rangle - 4t' |38\rangle - t |43\rangle - t |44\rangle + 2t' |50\rangle - 2t' |51\rangle - t |65\rangle + t |66\rangle
\nonumber\\
&+& 4t' |69\rangle - 4t' |70\rangle,
\nonumber\\
\hat H |53\rangle &=& - 2t |8\rangle + 2t |12\rangle + 4t' |55\rangle - 4t' |57\rangle + 2t |65\rangle,
\nonumber\\
\hat H |54\rangle &=& 2t |10\rangle - 2t |13\rangle + 4t' |56\rangle + 4t' |57\rangle + 2t |66\rangle,
\nonumber\\
\hat H |55\rangle &=& t |9\rangle + t |12\rangle + t |14\rangle + t |16\rangle - t |27\rangle - t |28\rangle + t |42\rangle + 
t |45\rangle + t |47\rangle 
\nonumber\\
&-& t |48\rangle + 4t' |53\rangle - 4t' |58\rangle - 4t' |61\rangle + 4t' |63\rangle + t |65\rangle + t |67\rangle,
\nonumber\\
\hat H |56\rangle &=& - t |11\rangle - t |13\rangle - t |15\rangle - t |16\rangle + t |26\rangle + t |28\rangle + t |41\rangle + 
t |46\rangle + t |47\rangle 
\nonumber\\
&+& t |49\rangle + 4t' |54\rangle + 4t' |58\rangle + 4t' |62\rangle + 4t' |63\rangle + t |66\rangle + t |67\rangle,
\nonumber\\
\hat H |57\rangle &=& t |8\rangle + t |10\rangle + t |14\rangle + t |15\rangle - t |26\rangle - t |27\rangle + t |41\rangle - 
t |42\rangle - t |43\rangle 
\nonumber\\
&-& t |44\rangle - 4t' |53\rangle + 4t' |54\rangle - 4t' |61\rangle - 4t' |62\rangle - t |65\rangle + t |66\rangle,
\nonumber
\end{eqnarray}

\newpage

\begin{eqnarray}
\hat H |58\rangle &=& - 2t |16\rangle - 2t |45\rangle + 2t |46\rangle - 4t' |55\rangle + 4t' |56\rangle,
\nonumber\\
\hat H |59\rangle &=& 4t' |23\rangle - 4t' |25\rangle + 2t |44\rangle - 2t |48\rangle + 2t |65\rangle,
\nonumber\\
\hat H |60\rangle &=& - 4t' |24\rangle + 4t' |25\rangle - 2t |43\rangle + 2t |49\rangle + 2t |66\rangle,
\nonumber\\
\hat H |61\rangle &=& - 2t |14\rangle + 2t |44\rangle + 2t |48\rangle - 4t' |55\rangle - 4t' |57\rangle,
\nonumber\\
\hat H |62\rangle &=& - 2t |15\rangle + 2t |43\rangle + 2t |49\rangle + 4t' |56\rangle - 4t' |57\rangle,
\nonumber\\
\hat H |63\rangle &=& 2t |9\rangle - 2t |11\rangle + 4t' |55\rangle + 4t' |56\rangle + 2t |67\rangle,
\nonumber\\
\hat H |64\rangle &=& 4t' |23\rangle - 4t' |24\rangle + 2t |45\rangle + 2t |46\rangle + 2t |67\rangle,
\nonumber\\
\hat H |65\rangle &=& - 2t' |8\rangle + 2t' |12\rangle - 2t' |15\rangle + 2t' |16\rangle + t |20\rangle - t |22\rangle + 
t |23\rangle - t |25\rangle + 2t' |44\rangle 
\nonumber\\
&-& 2t' |48\rangle + t |50\rangle - t |52\rangle + 2t |53\rangle + t |55\rangle - t |57\rangle + 2t |59\rangle - 2t' |66\rangle + 
2t' |67\rangle 
\nonumber\\
&-& 2t |68\rangle + 2t |70\rangle,
\nonumber\\
\hat H |66\rangle &=& 2t' |10\rangle - 2t' |13\rangle + 2t' |14\rangle - 2t' |16\rangle - t |21\rangle + t |22\rangle - t |24\rangle + t |25\rangle - 2t' |43\rangle 
\nonumber\\
&+& 2t' |49\rangle + t |51\rangle + t |52\rangle + 2t |54\rangle + t |56\rangle + t |57\rangle + 2t |60\rangle - 2t' |65\rangle + 
2t' |67\rangle 
\nonumber\\
&+& 2t |68\rangle + 2t |69\rangle,
\nonumber\\
\hat H |67\rangle &=& 2t' |9\rangle - 2t' |11\rangle + 2t' |14\rangle - 2t' |15\rangle + t |20\rangle - t |21\rangle + t |23\rangle - t |24\rangle + 2t' |45\rangle 
\nonumber\\
&+& 2t' |46\rangle + t |50\rangle + t |51\rangle + t |55\rangle + t |56\rangle + 2t |63\rangle + 2t |64\rangle + 2t' |65\rangle + 
2t' |66\rangle 
\nonumber\\
&+& 2t |69\rangle + 2t |70\rangle,
\nonumber\\
\hat H |68\rangle &=& - 2t |16\rangle + 4t' |22\rangle - 4t' |50\rangle + 4t' |51\rangle - 2t |65\rangle + 2t |66\rangle,
\nonumber\\
\hat H |69\rangle &=& 2t |14\rangle - 4t' |21\rangle + 4t' |50\rangle + 4t' |52\rangle + 2t |66\rangle + 2t |67\rangle,
\nonumber\\
\hat H |70\rangle &=& - 2t |15\rangle + 4t' |20\rangle + 4t' |51\rangle - 4t' |52\rangle + 2t |65\rangle + 2t |67\rangle.
\label{A1}
\end{eqnarray}

\newpage 

\section{The construction of the wave vectors $|i\rangle$}

In this appendix we present the construction of the wave vectors $|i\rangle$
presented in Figs.3-8 and used in Eq.(\ref{A1}). Each vector $|i\rangle$
has maximum 8 components and can be written as
\begin{eqnarray}
|i\rangle= N_i \sum_{m=1}^8 |i_m\rangle,
\label{B1}
\end{eqnarray}
where $N_i$ is a numerical factor preserving the normalization to unity, and
$|i_m\rangle$ represents the mathematical expression based on 
Eqs.(\ref{Eq2},\ref{Eq3},\ref{Eq4}) of the plotted particle configurations
${\cal{C}}_{i,m}$, $m=1,2,..8$ presented in the row $|i\rangle$ of Figs.3-8.
If the row $|i\rangle$ from Figs.3-8 contains less than 8 contributions, that 
means that some of $|i_m\rangle$ components taken at fixed $i$ coincide. Note
that in a fixed row $|i\rangle$ of Figs.3-8, different contributions are 
plotted in the order of increasing $m$ index. 

If at fixed $i$, the $m=1$ local particle configuration ${\cal{C}}_{i,1}$
is known (this is plotted in the first position of the row $|i\rangle$), 
all local particle configurations ${\cal{C}}_{i,m}$,
$m=2,3,..8$ can be deduced from it as follows: 
One takes the four axes defined by
$\gamma=x,y_1,y_2,a$ in Fig.20, and define the transformations:
$Tr(\gamma \ne x)$ as the translation (in the axis direction) along the axis 
$\gamma \ne x$ by vector ${\bf b}$ whose length is equal to the distance to the
nearest neighbor along the axis; and $R(\gamma \ne a)$ as a rotation with $\pi$
along the axis $\gamma$.  

\begin{figure} [h]                                                          %
\centerline{\includegraphics[width=4.6cm,height=5.3cm]{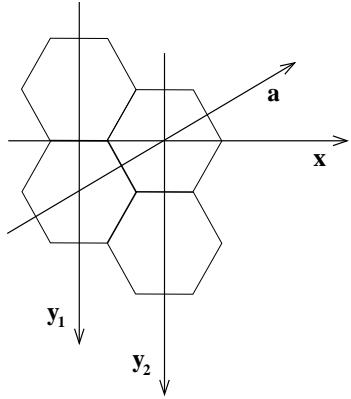}} %
\caption{The axes $\gamma=x,y_1,y_2,a$ of the transformations leading to     %
the components $|i_m\rangle$ at fixed $i$.}                                 %
\end{figure}                                                                %

With these conventions, for all fixed $i$ values, ${\cal{C}}_{i,m >1}$ can be 
obtained as
\begin{eqnarray}
&&{\cal{C}}_{i,2} = Tr(y_1) {\cal{C}}_{i,1}, \quad
{\cal{C}}_{i,3} = R(y_1) {\cal{C}}_{i,1}, \quad
{\cal{C}}_{i,4} = Tr(y_1) {\cal{C}}_{i,3}, \quad
{\cal{C}}_{i,5} = [R(x)Tr(a)] {\cal{C}}_{i,1},
\nonumber\\
&&{\cal{C}}_{i,6} = Tr(y_2) {\cal{C}}_{i,5}, \quad
{\cal{C}}_{i,7} = R(y_2) {\cal{C}}_{i,5}, \quad
{\cal{C}}_{i,8} = Tr(y_2) {\cal{C}}_{i,7}.
\label{B2}
\end{eqnarray}

For exemplification, Fig.21 presents the deduction procedure of the 
$|i_m\rangle$ components for $i=8$ and $i=31$.
 
\begin{figure} [h]                                                         %
\centerline{\includegraphics[width=15cm,height=10cm]{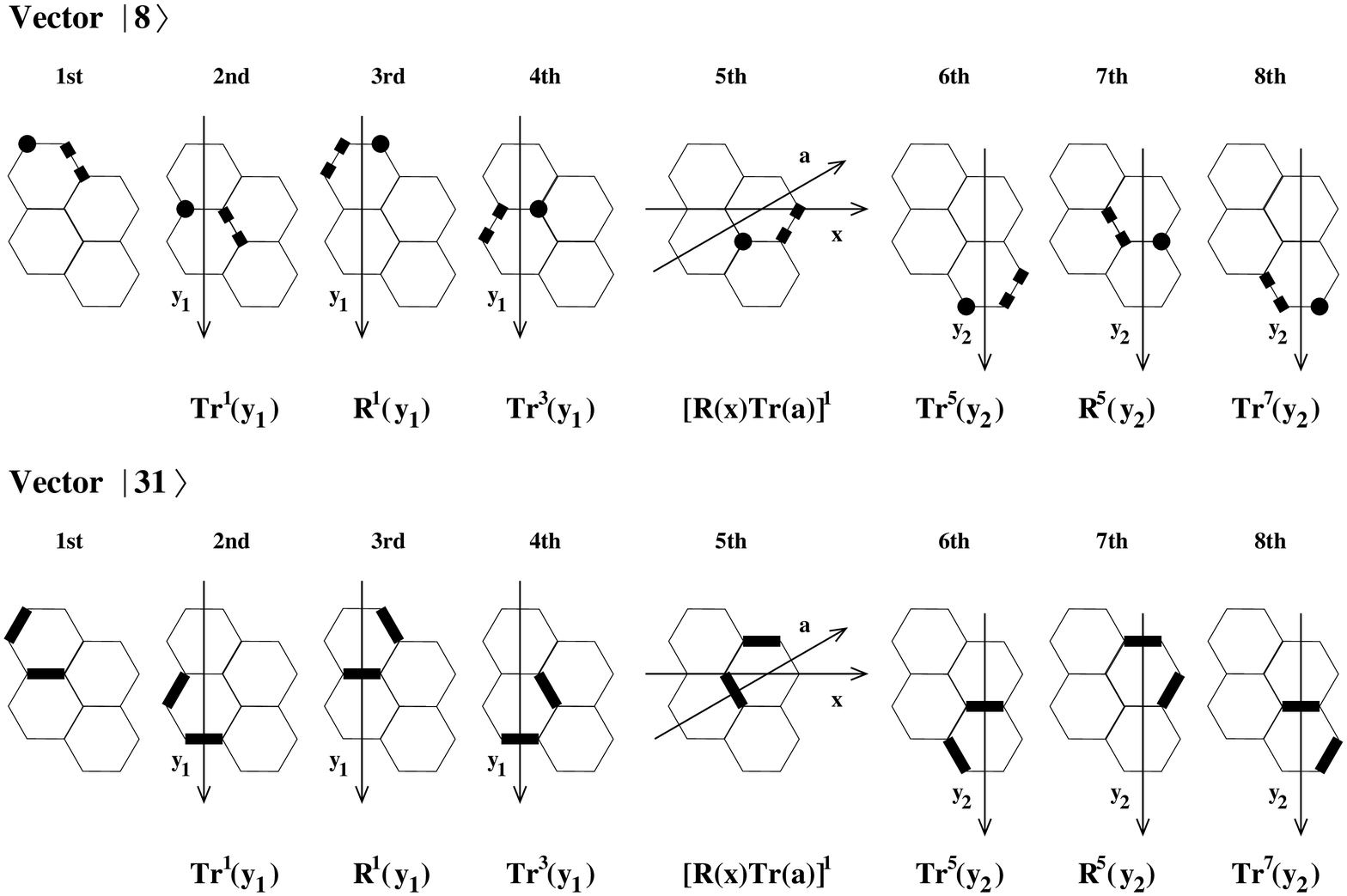}} %
\caption{The transformation leading to different components of the         %
vectors $|8\rangle$ and $|31\rangle$. For $P=Tr,R$, the notation           %
${\cal{C}}_{i,m_1}=P(\gamma) {\cal{C}}_{i,m_2}=P^{m_2}(\gamma)$ is used in the  %
plot, where $P^{m_2}(\gamma)$ is indicated below, while $m_1$ above the      %
figure.}                                                                   %
\end{figure}                                                               %


\begin{references}
\bibitem{I5}
One notes that in the Appendix A, Eq.(A1) of Ref.\cite{I1}, four misprint
have been observed, namely: a) in the right side of the equation for 
$\hat H |2\rangle$, instead of $|1\rangle$, $4|1\rangle$ must be written, b) in
the right side of the equation for $\hat H |3\rangle$, instead of 
$-2|3\rangle$, $-2|13\rangle$ must be written, c) in the right side of the
equation for $\hat H |21\rangle$, instead of $|43\rangle$, $-|43\rangle$ must 
be written, d) in the right side of the equation for $\hat H |35\rangle$, 
instead of $-|75\rangle$, $+|75\rangle$ must be written.
\bibitem{expl1}
The order of magnitude differences between Figs.14-15 can be attributed to
the different number of components in the vector $|1\rangle$ vector in the
hexagonal and square system cases.
\end{references}
\end{document}